\documentclass[12pt, a4paper]{article}
\usepackage[dvipdfmx]{graphicx}
\usepackage{amssymb}
\usepackage{amsmath}
\usepackage{bm}
\usepackage{color}
\usepackage{cite}
\usepackage{slashed}
\usepackage{subfigure}
\usepackage{epstopdf}            
\usepackage{epsfig}
\usepackage{here}
            
\renewcommand{\thefootnote}{\fnsymbol{footnote}}
\numberwithin{equation}{section} % Eq.(Sec.eq.)
\def\beq#1\eeq{\begin{align}#1\end{align}}

\renewcommand{\arraystretch}{1.3}

\RequirePackage{xspace}
\def\Bbar    {\kern 0.18em\overline{\kern -0.18em B}{}\xspace}
\def\Bb      {\ensuremath{\Bbar}\xspace}

\usepackage{caption}
\definecolor{BlueViolet}{rgb}{0.2, 0.00, 0.7}
\definecolor{Blue}{rgb}{0.15, 0.00, 0.9}
\usepackage[%dvipdfmx,
colorlinks=true, linkcolor=Blue,citecolor=Blue,urlcolor=BlueViolet]{hyperref}

\allowdisplaybreaks[4]
\def\beq#1\eeq{\begin{align}#1\end{align}}
\newcommand{\beqa}{\begin{eqnarray}}
\newcommand{\eeqa}{\end{eqnarray}}

\newcommand{\bpm}{\begin{pmatrix}}
\newcommand{\epm}{\end{pmatrix}}

\newcommand{\splitFunc}[1]{\mathrm{Split}}

\newcommand{\ctext}[1]{\raise0.15ex\hbox{\textcircled{\scriptsize{#1}}}}

\makeatletter
\def\EE{\@ifnextchar-{\@@EE}{\@EE}}
\def\@EE#1{\ifnum#1=1 \times10 \else \times10^{#1}\fi}
\def\@@EE#1#2{\times10^{-#2}}
\makeatother

\newcommand\unit[1]{\,\,\mathrm{#1}}
\newcommand\iab{\unit{ab^{-1}}}
\newcommand\ifb{\unit{fb^{-1}}}

\def\Babar{{\mbox{\slshape B\kern-0.1em{\smaller A}\kern-0.1em B\kern-0.1em{\smaller A\kern-0.2em R}}}}

\newcommand{\nubar}{\overline{\nu}}

\RequirePackage{xspace}
\def\Bbar    {\kern 0.18em\overline{\kern -0.18em B}{}\xspace}
\def\Bb      {\ensuremath{\Bbar}\xspace}

\begin{document}
\begin{titlepage}
% bold applies to math too
\begin{center}
\hfill{P3H-22-010, TTP22-004}
\vspace{.2 in}

{\Large\bf Revival of $H^-$ interpretation of $R_{D^{(*)}}$ anomaly and closing low mass window}\\
\vspace{.2 in}
{\large{Syuhei Iguro$^{a,b}$}}
\\
\vspace{.2 in}
$^{\rm (a)}${\it Institute for Theoretical Particle Physics (TTP), Karlsruhe Institute of Technology (KIT),
Engesserstra{\ss}e 7, 76131 Karlsruhe, Germany}\\
$^{\rm (b)}${\it Institute for Astroparticle Physics (IAP),
Karlsruhe Institute of Technology (KIT), 
Hermann-von-Helmholtz-Platz 1, 76344 Eggenstein-Leopoldshafen, Germany}\\
\vspace{.1 in}
{igurosyuhei@gmail.com}

\end{center}
%\vskip .3in

\begin{abstract}
%\si{memo: I checked the numerical result at least three times. the date of the last check is 12/01/2022.
%I checked the manuscript four times. The date of the last check was 16/01/2022}\\
Thanks to the recent careful revisit of the theoretical prediction of the $B_c$ meson lifetime, the conservative upper bound on the branching ratio (BR) of $\tau \nu$ mode is found to be $\simeq 63\%$ due to the large charm quark mass uncertainty. 
Although it is well known that a charged Higgs ($H^-$) interpretation of the $R_{D^{(*)}}$ anomaly is excluded by the previously proposed bounds, BR$(B_c\to\tau\nu)\le 30\%$ and $\le10\%$, $H^-$ can still explain the anomaly within $1\sigma$ if we adopt the 63$\%$ one. 
The scalar contribution is also favored by the polarization data $F_L^{D^*}$ measured at the Belle.
Since the implied NP scale is within the reach of the Large Hadron Collider (LHC), collider searches are powerful tools to test the scenario.
For instance, the $\tau\nu$ resonance search has already put the more stringent bound for $m_{H^-}\ge 400$\,GeV.
In this work we revisit the further lighter mass range, $180\,$GeV$\le m_{H^-}\le 400\,$GeV which has not been covered yet.
We will see that a combination of the conventional stau search and low mass flavor inclusive and bottom flavored di-jet resonance searches can place a new limit on the interpretation.
We summarize the current status of the low mass region and discuss the future sensitivity in the high luminosity (HL)-LHC based on the existent collider constraints. 
\end{abstract} 
%end of abstract

\end{titlepage}
\renewcommand{\thefootnote}{\#\arabic{footnote}}
\setcounter{footnote}{0}

%%%%%%%%%%%%%%%%%%%%%%%%%%%%%%%%%%%%%%%%%%%%%%%%%%%
\section{\boldmath Introduction}
\label{Sec:introduction}
%%%%%%%%%%%%%%%%%%%%%%%%%%%%%%%%%%%%%%%%%%%%%%%%%%%

%%%%% status of RD, RD* %%%%%
The lepton flavor universality (LFU) is one of the most important predictions  within the standard model (SM) and thus if the violation is observed, it immediately implies the existence of the physics beyond the SM.
The $R_{D^{(\ast)}}$ discrepancy reported by B-factories \cite{Lees:2012xj,Lees:2013uzd,Huschle:2015rga,Sato:2016svk,Hirose:2016wfn,Abdesselam:2019dgh,Aaij:2015yra,Aaij:2017deq}, where 
$R_{D^{(\ast)}}=\text{BR}(\overline{B}\to D^{(*)}\tau\nubar)/\text{BR}(\overline{B}\to D^{(*)}\ell\nubar)$,
with $\ell=\mu$ for LHCb and an average of $e$ and $\mu$ for BaBar and Belle is defined, suggests violation of the LFU between $\tau$ and light leptons.
The current significance of the deviation is about 3-4$\sigma$ \cite{Aoki:2021kgd,Iguro:2020cpg} and it would be natural to think the extension of Higgs sector of the SM since we have the mass hierarchy in leptons.  
%%%%% G2HDM %%%%%
A generic two Higgs doublet model (G2HDM) where an additional Higgs doublet with couplings to all fermions is added is one of the simplest extensions of the SM which often appears in a UV theory e.g. a left-right symmetric model \cite{Iguro:2018oou,Iguro:2021nhf}.
In the G2HDM there are 4 additional degrees of freedom, a CP even scalar ($H$), a CP odd scalar ($A$) and charged scalars ($H^\pm$).
Such an extension, however, can be dangerous since the additional scalars have flavor violating interactions even at tree level in general, the model had been attracting attentions in light of the discrepancy \cite{Crivellin:2012ye,Celis:2012dk,Crivellin:2013wna,Cline:2015lqp,Crivellin:2015hha,Celis:2016azn,Lee:2017kbi,Iguro:2017ysu,Iguro:2018qzf,Martinez:2018ynq,Fraser:2018aqj,Cardozo:2020uol,Athron:2021jyv} since $H^-$ can contribute to $\overline{B}\to D^{(*)}\tau\nu$ process.
The charged Higgs effect can be generally encoded in the low-energy effective Hamiltonian, 
\begin{align}
 {\mathcal H}_{\rm{eff}}= 
 2 \sqrt 2 G_F V_{cb} \Bigl[ 
 & (\overline{c} \gamma^\mu P_Lb)(\overline{\tau} \gamma_\mu P_L \nu_{\tau}) +C_{S_R} (\overline{c} P_Rb)(\overline{\tau} P_L \nu_{\tau}) +C_{S_L}(\overline{c} P_L b)(\overline{\tau} P_L \nu_{\tau}) \Bigl{]},\label{Eq:Hamiltonian}
\end{align} 
with $P_{L/R}=(1\mp\gamma_5)/2$.
In this paper, right-handed neutrinos are not considered.\footnote{See, Ref.~\cite{Iguro:2018qzf} for a model and the analysis with light right-handed neutrinos in the context of the $R_{D^{(\ast)}}$ anomaly.}
Here, the Wilson coefficients (WCs) are normalized by the SM contribution as, 
$\mathcal{H}_{\rm eff} = 2 \sqrt 2 G_F V_{cb} (\overline{c} \gamma^\mu P_Lb)(\overline{\tau} \gamma_\mu P_L \nu_{\tau})$, corresponding to $C_{S_{L,R}}=0$.
Note that the SM contribution is suppressed by the Cabibbo-Kobayashi-Maskawa (CKM) matrix element $V_{cb}$, where $V_{cb}=0.042$ is fixed throughout this paper corresponding to the inclusive $V_{cb}$~\cite{Zyla:2020zbs}. 

It is well known that the $B_c$ meson lifetime constrains the $H^-$ interpretation.
Within the SM the branching ratio of the $B_c\to l\nu$ decay, which is described by the same Hamiltonian contributing to $\bar{B}\to D^{(*)} l\bar{\nu}$ is suppressed by the final lepton mass to flip the chirality.
On the other hand, the contributions with scalar operators are not suppressed and easily enhance the decay branching ratio of $B_c\to\tau\nu$ when one want to enhance BR$(\bar{B}\to D^{(*)} \tau\bar{\nu})$.

%%%%% Brief History of the Bc bound  %%%%%%
In 2016, Ref.\,\cite{Alonso:2016oyd} derived BR$(B_c\to\tau\nu)\le 30\%$  based on the $B_c$ lifetime \cite{Beneke:1996xe} and BR$(B_c\to\tau\nu)\le 10\%$ based on the LEP data
is suggested in 2017 \cite{Akeroyd:2017mhr}.
However, the underestimation of the charm mass uncertainty and the scale dependence of the $b\to B_c$ fragmentation function are pointed out and the conservative bound is estimated to be BR$(B_c\to\tau\nu)\lesssim 60\%$ \cite{Blanke:2018yud}.
The recent careful revisit gives the more conservative bound of BR$(B_c\to\tau\nu)\lesssim 63\%$ \cite{Aebischer:2021ilm}.

According to the relaxed constraint from $B_c\to\tau\nu$ and the previous experimental result from the Belle experiment in 2019 \cite{Belle:2019rba} which favors the more SM like $R_{D^{(*)}}$ with reduced uncertainties, the scalar interpretation has silently revived.
It is noted that the scalar contribution is also favored by the $D^*$ polarization, $F_L^{D^*}$ reported by Belle \cite{Belle:2019ewo} which is observed to be slightly larger than the SM prediction.
Future data may prefer the more SM like $R_{D^{(*)}}$ with reduced uncertainty, and hence it is always important to clarify the range of the possible enhancement in each model.

Since the implied NP scale is within the reach of the LHC, it is interesting to study the LHC sensitivity for the scenarios.
Ref.\,\cite{Iguro:2018fni} used the existent CMS result with 36$\ifb$ of the data at $\sqrt{s}=13\,$TeV, which searches for the high mass $\tau\nu$ resonance motivated by $W^\prime$ in a sequential standard model \cite{Sirunyan:2018lbg} to constrain the $H^-$ explanation.
The experimental upper limit on signal events number is available for $m_{W^\prime}\ge400$\,GeV.
It has resulted in the exclusion of the $1\sigma$ interpretation at the time for $m_{H^-}\ge400$\,GeV through $pp\to bc\to\tau\nu$ process.
The data for $m_{H^-}\le400$\,GeV is not available in Ref.\,\cite{Sirunyan:2018lbg} since the lighter resonance search is suffered from the huge SM background (BG) from W boson and the original motivation is to push up the lower limit for heavy $W^\prime$.
Although the result at $\sqrt{s}=8\,$TeV was also available from $m_{W^\prime}\ge300\,$GeV \cite{CMS:2015hmx}, its constraint was not studied well since the primary goal of the paper was to set the stringent bound for heavy scenarios \cite{Iguro:2018fni}.  

In this work we revisit the low mass $H^-$ interpretation with available collider constraints.
We will see that a combination of the low mass flavor inclusive and bottom flavored di-jet resonance searches \cite{CMS:2017dcz,CMS:2018kcg,ATLAS:2019itm} and conventional stau search \cite{CMS:2021woq} allows us to probe the wide range of the remaining parameters of a low mass $H^-$ scenario.

%%%%%  menu %%%%%%
This paper is organized as follows.
A model setup and the current status of the $H^-$ interpretation of the $R_{D^{(*)}}$ anomaly are explained in Sec.~\ref{Sec:model}.
There we also discuss the collider constraint and
impact on the $H^-$ interpretation.
Sec.~\ref{Sec:summary} is devoted to conclusions and discussion.
The main text is supported by the appendix discussing box induced $H^-$ contribution to B meson mixings and providing additional figures.

%%%%%%%%%%%%%%%%%%%%%%%%%%%%%%%%%%%%%%%%%%%%%%%%%%%
\section{Current status of the $H^-$ interpretation}
\label{Sec:model}
%%%%%%%%%%%%%%%%%%%%%%%%%%%%%%%%%%%%%%%%%%%%%%%%%%%
In this section, we introduce the simplified model of a charged Higgs based on a general two Higgs doublet model \cite{Iguro:2017ysu} and discuss the current status of the $H^-$ interpretation of the anomaly.
%%%%%%%%%%%%%%%%%%%%%%%%%%%%%%%%%%%%%%%
\subsection{Current status of the scalar operator} 
%%%%%%%%%%%%%%%%%%%%%%%%%%%%%%%%%%%%%%%
Before discussing the model dependent constraint, let us summarize the model independent status of the scalar interpretation of the anomaly based on the weak effective field theory.
It is known that the right handed quark scalar current can not explain the anomaly, we will focus on the case where $C_{S_L}\neq0$.
Assuming the real WC scalar operator can not explain the discrepancy, however, complex WC which corresponds to complex Yukawa couplings can enhance $R_{D^{(*)}}$ and provide a good fit \cite{Iguro:2017ysu,Iguro:2018vqb}

As for the numerical descriptions of $R_D$, $R_{D^*}$, $F_L^{D^*}$, BR($B_c\to\tau\nu$) we follow \cite{Blanke:2018yud},
%%%%%%%%%%%%%%%%%%%%%%%%%%
\begin{align}
R_D&\simeq R_D^{SM}\biggl{(}1+1.54{\rm Re}\bigl[ C _{S_L}\bigl]+1.09|C _{S_L}|^2\biggl{)},\\
R_{D^*}&\simeq R_{D^*}^{SM}\biggl{(}1-0.13{\rm Re}\bigl[ C _{S_L}]+0.05|C _{S_L}|^2\biggl{)},\\
F_L^{D^*}&\simeq 
(0.46-0.13{\rm Re}\bigl[ C _{S_L}]+0.05|C _{S_L}|^2)/(1-0.13{\rm Re}\bigl[ C _{S_L}]+0.05|C _{S_L}|^2),\\
\rm{BR}&(B_c\to\tau\nu)\simeq 0.02|1-4.3C_{S_L}|^2.
\label{Eq:RDS}
\end{align}
%%%%%%%%%%%%%%%%%%%%%%%%%%
Here the WC is defined at $m_b=4.2$\,GeV.
Similar numerical formulae can be found in Ref.\,\cite{Iguro:2018vqb}.

Fig.\,\ref{Fig:current_EFT} shows the current status of the scalar contribution.
The experimental result is shown in red ellipsis. 
The SM prediction denoted in a yellow star is taken from the HFLAV2021 \cite{Aoki:2021kgd}.
Varying $C_{S_L}$ in the complex plane uniquely gives the prediction on the plane.
The grey shaded region is out of the prediction with $C_{S_L}$ and blue and magenta lines show the prediction for $F_L^{D^{(*)}}$ and BR$(B_c\to\tau\nu)$.
If we adopt the BR$(B_c\to\tau\nu)\le63\%$ bound, the region above the magenta solid line is excluded.
In that case the scalar operator can still explain the anomaly within $1\sigma$ and also enhances $F_L^{D^{(*)}}$ up to 0.54 which comes closer to the experimental value $F_L^{D^{(*)}}=0.60\pm0.09$ \cite{Belle:2019ewo}.
It is worth noting that only scalar contributions can enhance $F_L^{D^{(*)}}$. 
%%%%%%%%%%%%%%%%%%%%%%%%%%%%%%%%%%%%%%%%%%%%%%%%%
\begin{figure}[t]
\begin{center}
\includegraphics[width=28em]{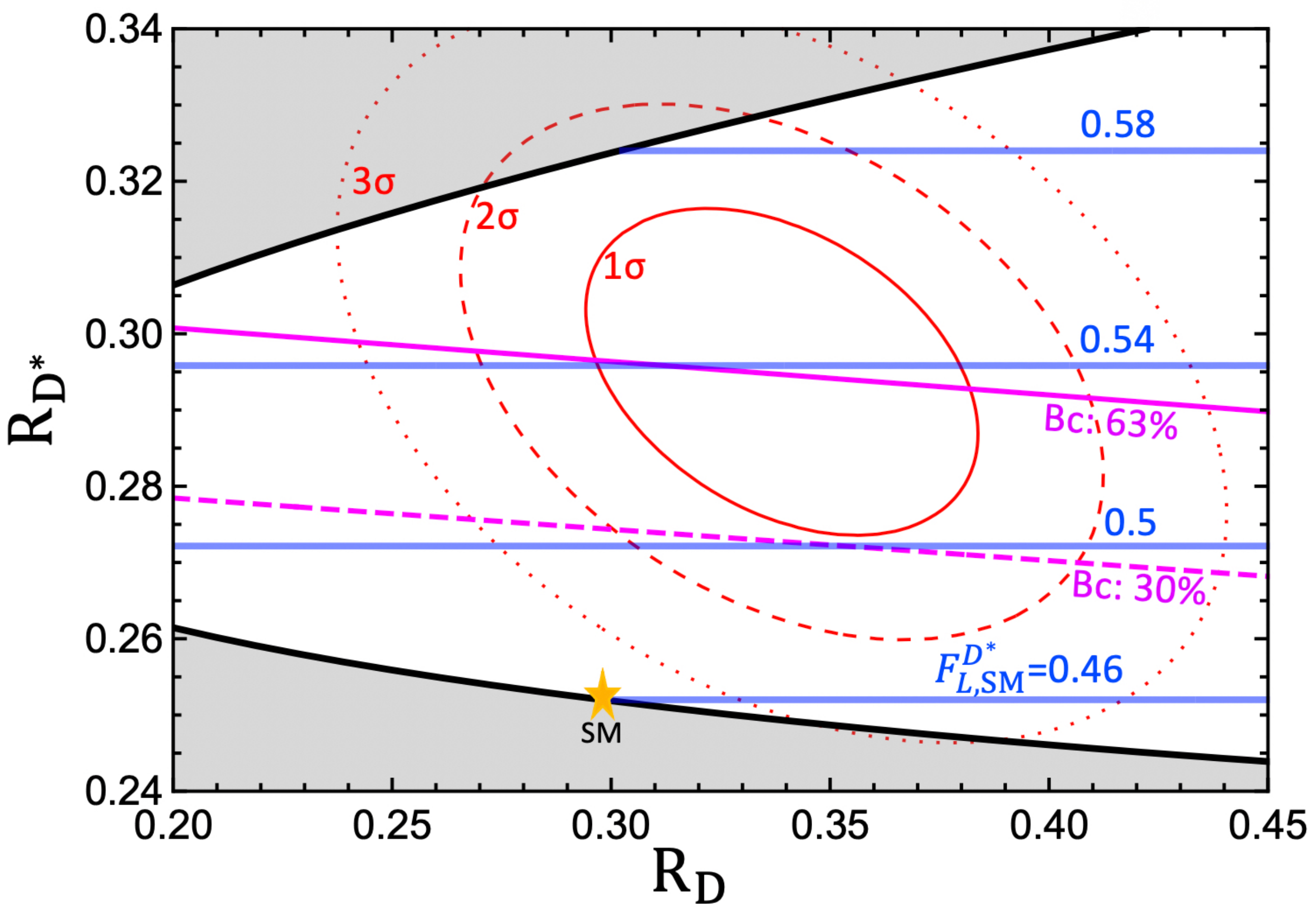} 
\caption{
\label{Fig:current_EFT}
The current status of the scalar interpretation of the $R_{D^{(*)}}$ anomaly.
The world average of the data at 1,\, 2 and 3\,$\sigma$ are shown by the red solid, dashed and dotted ellipsis.
Horizontal blue lines show the correlation with $F_L^{D^*}$.
Horizontal magenta solid (dashed) line corresponds to BR($B_c\to\tau\nu)=63\,(30)\%$.
The SM prediction is shown in a yellow star taken from the HFLAV2021.
Grey shaded region is out of the model prediction.
} 
\end{center}
\end{figure}
%%%%%%%%%%%%%%%%%%%%%%%%%%%%%%%%%%%%%%%%%%%%%%%%%

%%%%%%%%%%%%%%%%%%%%%%%%%%%%%%%%%%%%%%%
\subsection{Model and mass range} 
%%%%%%%%%%%%%%%%%%%%%%%%%%%%%%%%%%%%%%%
The interaction Lagrangian of the heavy scalars relevant to $R_{D^{(*)}}$ in the Higgs basis is given as
\begin{align}
{\cal L}_{int}=&
+y_{Q_u} \frac{H+iA}{\sqrt{2}} (\overline{t} P_R c)
+ y_{Q_d}\frac{H-iA}{\sqrt{2}} (\overline{s} P_R b)
+ y_{\tau} \frac{H-iA}{\sqrt{2}} (\overline{\tau} P_R \tau)\nonumber\\
& + y_{Q_u} H^- (\overline{b} P_R c)
- y_{Q_d} H^- (\overline{b} P_L c) 
- y_\tau H^- (\overline{\tau} P_L \nu_{\tau})   +{\rm{h.c.}},\label{Eq:G2HDM}
\end{align}
where the neutral scalar interaction and the charged scalar interaction are related by the SU(2)$_
{\rm{L}}$ rotation.
Those Yukawa couplings are complex in general and provide complex WC in that case which is beyond the scope of the study performed in the literature \cite{Blanke:2018yud}.
Here the CKM suppressed terms such as $y_{Q_b}\frac{H-iA}{\sqrt{2}} (\overline{b} P_R b)$ are considered since it can not provide the large contribution and suffers from the direct search via $b\bar{b}\to \tau\bar{\tau}$ at the LHC \cite{Faroughy:2016osc}.

The alignment limit is taken and the SM Higgs couplings are the same as the original one.
With this coupling normalization $C_{S_L}=y_{Q_u}^* y_\tau/m_H^2/(2 \sqrt 2 G_F V_{cb})$ holds for instance.
It is noted that an upper bound on the mass is set by utilizing the $\tau\nu$ resonance search result by the CMS \cite{Sirunyan:2018lbg} with 36 fb$^{-1}$.
They report the upper limit on cross section (Xs) times BR for $m_{W^\prime}\ge 400$\,GeV.
Reinterpreting the bound based on the fast collider simulation excludes the interpretation at that time for $m_{H^-}\ge 400$\,GeV \cite{Iguro:2018fni}.
It is worth noting that the ATLAS with the Run 2 full data did not find a significant excess \cite{ATLAS:2021bjk}.
Hence it results in more stringent bound but they report the bound only for $m_{W^\prime}\ge 500\,$GeV. 
The Run 1 result is also available from $m_{W^\prime}=300$\,GeV, however, the constraint is weaker when one compares at $m_{W^\prime}=400$\,GeV \cite{CMS:2015hmx}.

Besides, the lower bound on the charged Higgs mass $m_{H^-}\ge 80$\,GeV is set by LEP experiment via the electroweak (EW) production $pp\to\gamma,\,Z\to H^- H^+$ which is followed by $H^-\to\tau\bar{\nu}$ \cite{ALEPH:2013htx}.
An EW precision observable, T parameter constrains the mass difference $|m_H-m_{H^-}|$ and/or $|m_A-m_{H^-}|$ \cite{Zyla:2020zbs}.
Therefore we assume the mass degeneracy among heavy scalars $m_H=m_A=m_{H^-}$ for simplicity.
In that case there could be constraints from the exotic top quark decay $t\to c \phi$, where $\phi$ is H and A induced by $y_{Q_u}$ defined in Eq.\,(\ref{Eq:G2HDM}) if the mass scale of the heavy scalar is sufficiently light.
Therefore we focus on the mass window 
\begin{align}
180\,\text{GeV}\le m_{H^-}\le 400\,\text{GeV},
\label{Eq:window}
\end{align}
which is currently not excluded by  collider and flavor constraints.

%%%%%%%%%%%%%%%%%%%%%%%%%%%%%%%%%%%%%%%
\subsection{Flavor constraint} 
%%%%%%%%%%%%%%%%%%%%%%%%%%%%%%%%%%%%%%%
Here we discuss the flavor constraints on the relevant Yukawa couplings.
In order to explain the $R_{D^{(*)}}$ anomaly the product $y_{Q_u}^*\times y_\tau $ and/or $y_{Q_d}^*\times y_\tau$ need to be sizable.
However, the quark Yukawa term of $y_{Q_d} H^- (\overline{c} P_R b)$ is stringently constrained by the neutral scalars mediated $B_s$-$\overline{B}_s$ mixing \cite{Branco:2011iw}. 
As a result, $C_{S_R}$ needs to be tiny and decouples from our discussion.
Therefore we set $y_{Q_d}=0$ and denote $y_{Q_u}$ as $y_{Q}$ for simplicity.
On the other hand the interaction of $y_{Q} H^- (\overline{c} P_L b)$ is less constrained since the SU(2)$_{\rm{L}}$ rotation leads to the interaction of $y_{Q} \phi (\overline{c} P_L t) $ which does not generate flavor violation among down quarks at tree level.
As a consequence, there are three relevant model parameters, $y_{Q}$, $y_\tau$ and $m_{H}$, and 
the relation $C_{S_L}=y_\tau y_{Q}^*/m_{H}^2 / ( 2 \sqrt 2 G_F V_{cb})$ holds at the heavy scalar scale.\footnote{We can discuss the other couplings like $y_t \phi (\bar{t}P_R t)$, however, its contribution to $C_{S_L}$ is small and $pp\to gg\to \phi\to \tau\bar{\tau} $ at the LHC constrains the size of $y_t$ stringently.
Consequently, it is not easy to drastically dilute the signal BR discussed bellow. 
See review-ish paper \cite{Iguro:2017ysu} and references therein for more quantitative discussion.}
This situation corresponds to Fig.\,\ref{Fig:current_EFT}.

In addition to $B_c\to\tau\nu$, 1-loop $H^-$ induced flavor processes e.g. $B$ mixings (box), $b\to s\gamma$ (penguin), $\epsilon_K$ (penguin) and $b\to s l\bar{l}$ are discussed in previous works \cite{Iguro:2017ysu,Iguro:2018qzf,Iguro:2019zlc}.
Among them, B meson mixings give the most stringent constraint on $y_{Q}$.
We adopt the constraint from Ref.\,\cite{DiLuzio:2019jyq}. 
The relevant expression of the $H^-$ contribution is given in Appendix\,\ref{Sec:App_flavor}.
On the other hand the constraint on $y_\tau$ via the vertex correction to the $Z\tau\bar{\tau}$ interaction is very weak and neglected \cite{Abe:2015oca}.
Besides 1-loop induced contribution to $g-2$ of $\tau$ is also small because of the absence of chirality enhancement which is often discussed in light of the muon $g-2$ anomaly , see Ref.\,\cite{Iguro:2019sly} for instance.
Furthermore the purely leptonic decay of the tau lepton does not change unless an additional Yukawa coupling to light lepton is considered which is not helpful for the $R_{D^{(*)}}$ discrepancy.
It is noted that the complex Yukawa couplings, $y_Q$ and $y_\tau$ do not induce contributions to the electron EDM even at two loop order in the alignment limit.

%%%%%%%%%%%%%%%%%%%%%%%%%%%%%%%%%%%%%%%
\subsection{Collider constraint on the low mass scenario}
\label{Sec:lowmass_collider}
%%%%%%%%%%%%%%%%%%%%%%%%%%%%%%%%%%%%%%%

%%%%%%%%%%%%%%%%%%%%%%%%%%%%%%%%%%%%%%%%%%%%%%%%%
\begin{figure}[t]
\begin{center}
\includegraphics[scale=0.5]{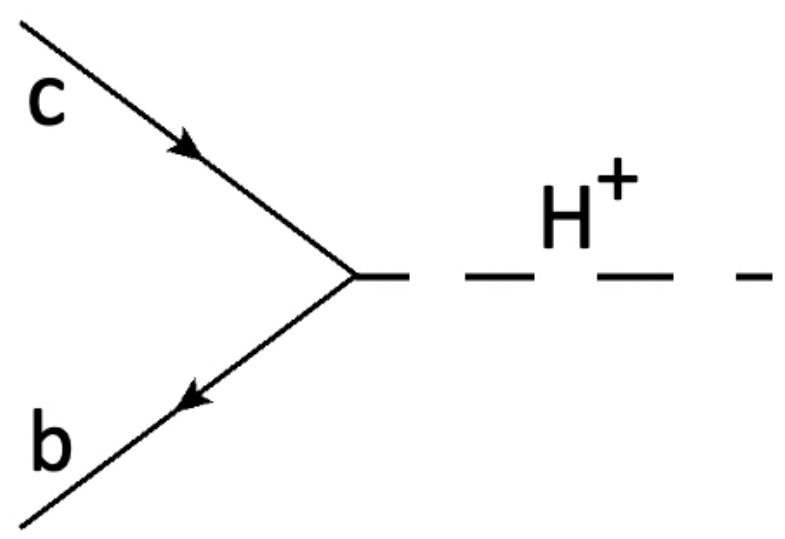}
\includegraphics[scale=0.5]{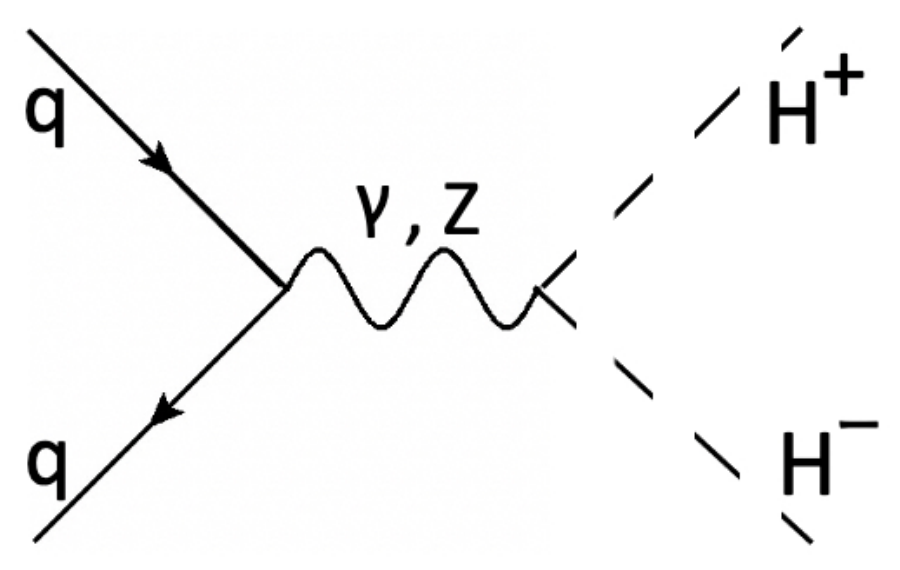} 
\includegraphics[scale=0.5]{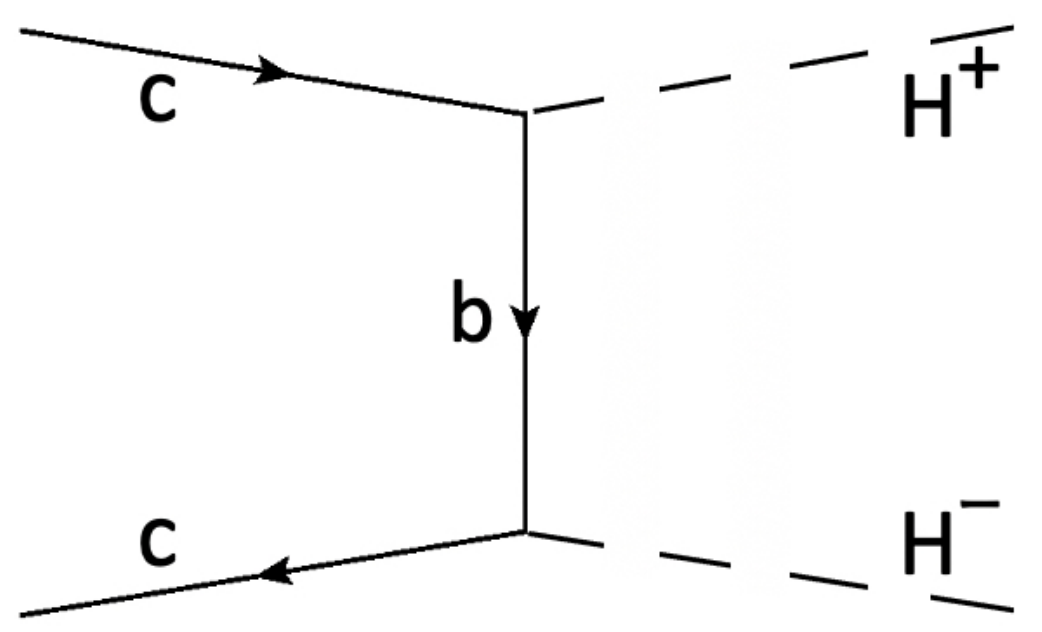} 
\caption{
\label{Fig:dia_Hp}
The representative diagrams for the single and pair production of charged Higgs are shown.
There is also a t-channel diagram where c and b are exchanged in the right panel.  
} 
\end{center}
\end{figure}
%%%%%%%%%%%%%%%%%%%%%%%%%%%%%%%%%%%%%%%%%%%%%%%%%

As mentioned above the orthodox $\tau\nu$ search constraint is not available in the full mass range of our interest.
In the presence of nonzero $y_Q$ and $y_\tau$, the charged scalar can decay into $\tau\nu$ and $bc$ while the neutral heavy scalars can decay into $tc$ and $\tau\bar{\tau}$.
The decay width of the $\tau\nu$ mode and $bc$ mode are expressed as
\begin{align}
    \Gamma(H^- \to \tau\bar{\nu})=\frac{|y_\tau|^2}{16\pi}m_H,~~~
    \Gamma(H^- \to b\bar{c})=\frac{3|y_Q|^2}{16\pi}m_H,
\end{align}
where fermion masses in the final state are neglected.  
It is noted that the $bc$ mode has a color factor.
The BR($H^-\to\tau\bar\nu$) and width to mass ratio on the $y_Q$ versus $y_\tau$ plane are shown in Fig.\,\ref{Fig:App_BR} of the Appendix \ref{Sec:App_other_mass}.
Since the $H^-$ width is smaller than $10\%$ of the mass in our case, the narrow width approximation is assumed. 
Although $y_Q$ can generate the same sign top signature mediated by neutral scalars, the mass degeneracy can suppress the amplitude \cite{Iguro:2018qzf}.
The mass degeneracy among heavy scalars is favored by T parameter, and thus the same sign top signature could not be a smoking gun signal of the model.

Single charged Higgs can be generated in a bc fusion and a pair of charged Higgs are produced via the EW production and t-channel b (c) quark exchange processes shown in Fig.\,\ref{Fig:dia_Hp}.
We derive the collider constraint from low mass bottom flavored di-jet search at $\sqrt{s}=8$ \cite{CMS:2018kcg}, flavor inclusive di-jet search at $\sqrt{s}=13$\,TeV \cite{CMS:2017dcz}, low mass bottom flavored di-jet with a high $p_T$ photon search \cite{ATLAS:2019itm} and stau search \cite{CMS:2021woq} with full run II data.
There are, however, many other results on di-jet resonances they are less stringent, looking for heavier particles and/or presenting the result in specific coupling planes \cite{ATLAS:2018hbc,CMS:2019emo,CMS:2019xai,CMS:2019mcu,ATLAS:2019fgd,ATLAS:2018tfk,ATLAS:2018qto,ATLAS:2020iwa,ATLAS:2021suo,ATLAS:2014ktg,CMS:2018pwl}.

The 8\,TeV bottom resonance result with 20\,fb$^{-1}$ of the data is available for the resonance mass heavier than $325\,$GeV and flavor blind result at $\sqrt{s}=13\,$TeV with 36\,fb$^{-1}$ of the data can constrain up to 300\,GeV.
The 13\,TeV bottom resonance with the photon result is available to put a bound for $225$\,GeV$\le m_{H^-}$.

Although they originally search for a bottom flavored di-jet resonance, the mistag rate $\slashed{c}\to b$ ($\epsilon_{c\to b}$) is not small and hence their result can be used to constrain the $bc$ resonance.
To keep the signal event number and reject the huge amount of QCD originated BG, Ref.\,\cite{CMS:2018kcg} required 2 bottom flavored jets, one passing the ``tight" selection and another passing the ``medium" selection.
The b-tagging efficiency of the ``tight" working point $\epsilon_{b\to b}$ is $50\%$ and QCD jet mistag rate $\epsilon_{j\to b}$ is 0.1$\%$.
On the other hand the efficiency of the ``medium" working point $\epsilon_{b\to b}$ is $70\%$ and the QCD jet mistag rate $\epsilon_{j\to b}$ is $1\sim2\%$.
The corresponding $c\to b$ mistag rates , however, are not explicitly written in Ref.\,\cite{CMS:2018kcg}, we can read them from Fig.\,6 of Ref.\,\cite{CMS:2012feb}, leading to $\epsilon_{c\to b}\simeq4\%$ for the ``tight" and $\epsilon_{c\to b}\simeq19\%$ for the ``medium" working points for Run\,1, respectively.\footnote{Rigorously speaking, the determination of the tagging efficiency is performed based on different processes from the processes of our interest.
The estimation of the correction factor to account for the event differences calls the detailed experimental analysis and is beyond the scope of the paper. 
Therefore the effect is neglected.}
The $c\to b$ mistag rates in the low mass bottom flavored di-jet with an additional high $p_T$ photon search is explicitly written in Ref.\,\cite{ATLAS:2019itm}.
The working point of $\epsilon_{b\to b}\simeq77\%$ and $\epsilon_{c\to b}\simeq 25\%$ was applied for both b jets tagging.

%%%%%%%%%%%%%%%%%%%%%%%%%%%%%%%%%%%%%%%%%%%%%%%%%
\begin{figure}[t]
\begin{center}
\includegraphics[scale=0.33]{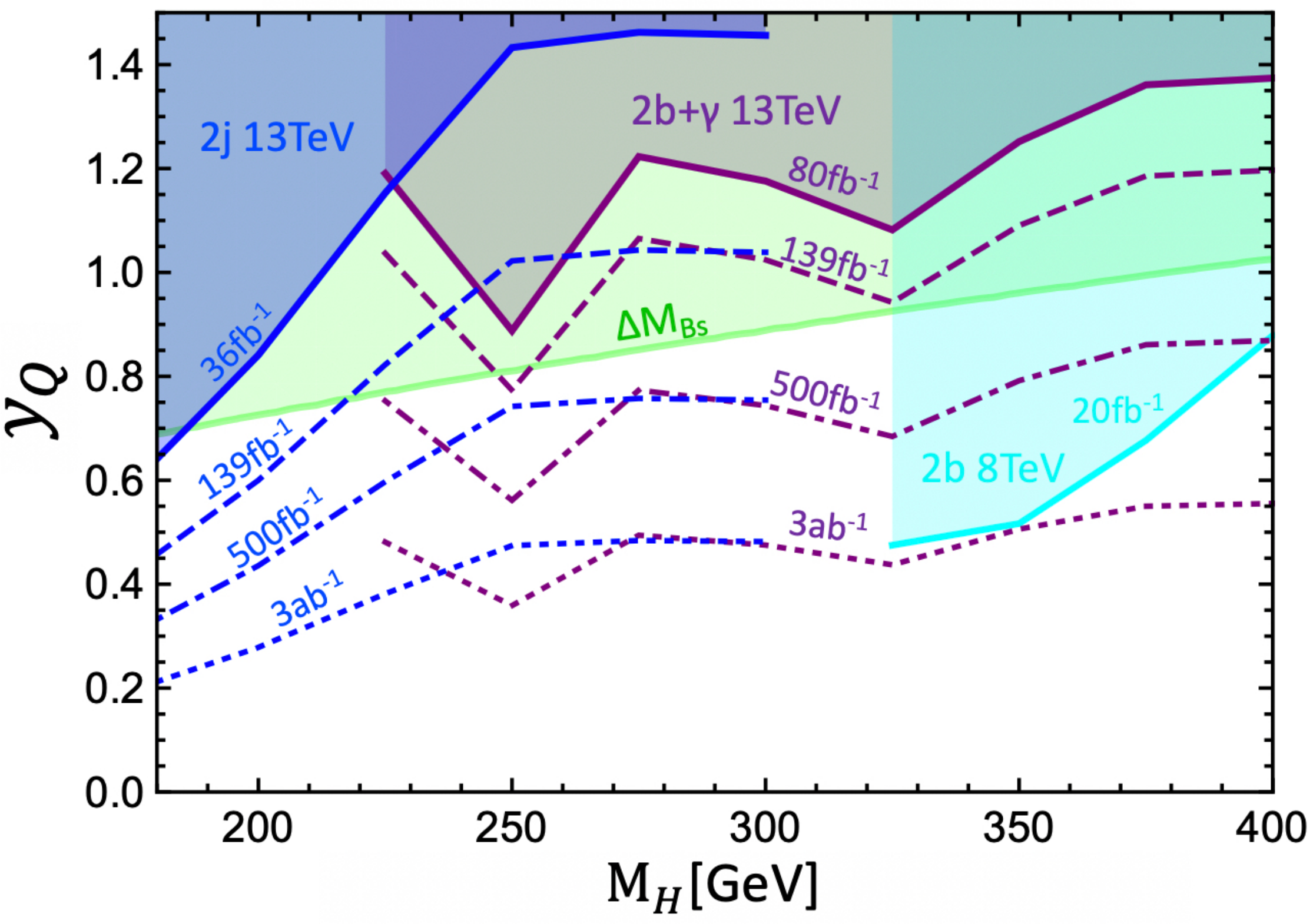}~
\includegraphics[scale=0.36]{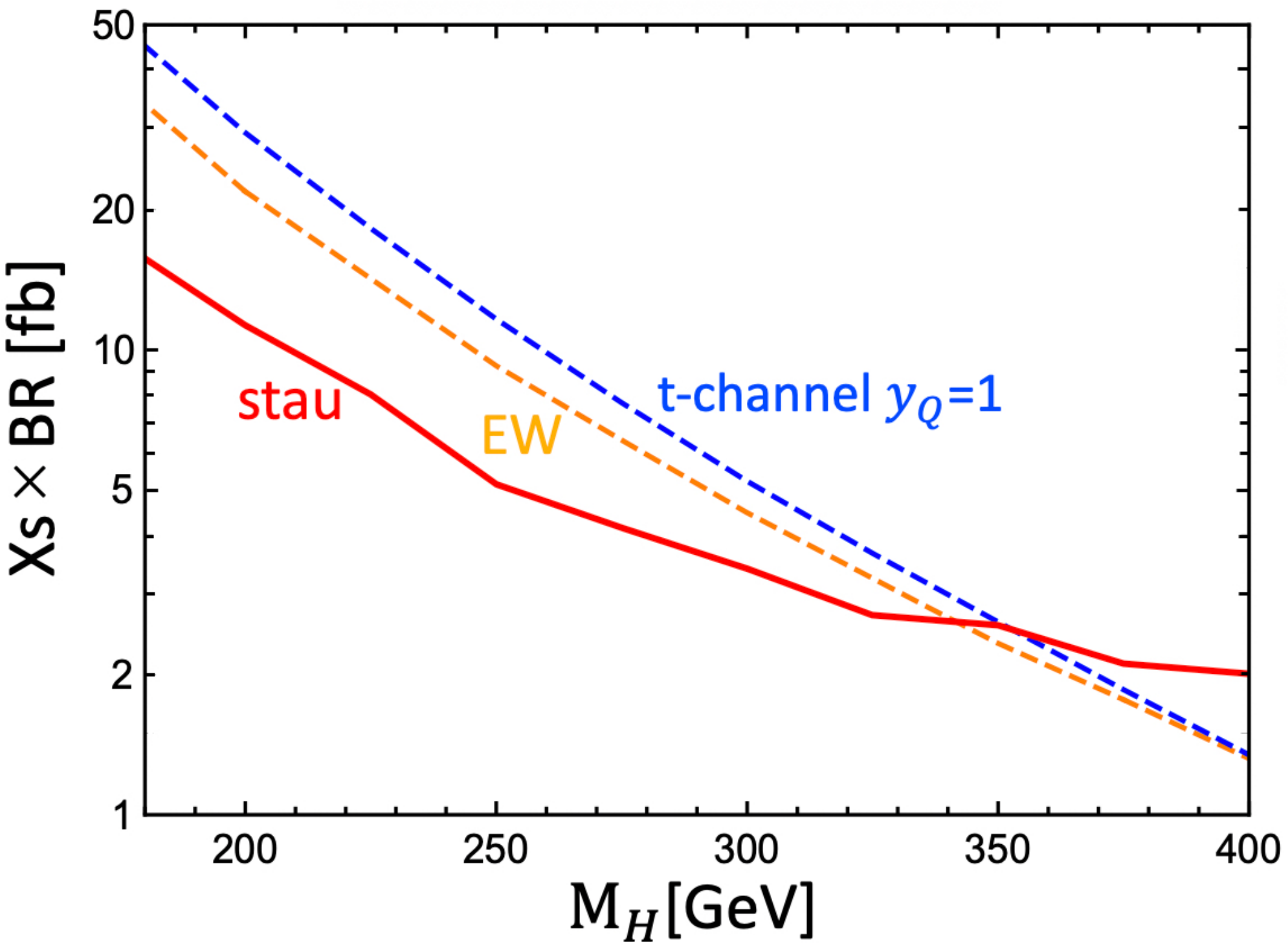}
\caption{
\label{Fig:yQvsmHp}
The $bc$ resonance constraints and B meson mixings constraint are shown in the mass versus $y_Q$ plane on the left.
Cyan, blue and purple shaded regions are excluded by $bc$ resonance based on the di-bottom flavored and flavor inclusive jet resonance search at $\sqrt{s}=8\,\rm{TeV}$ and $13\,\rm{TeV}$ and the di-bottom resonance with a high $p_T$ photon search at $13\,\rm{TeV}$, respectively.
The B meson mixings constraint is expressed in green.
In the right panel the production cross section of the EW pair production and $y_Q$ induced t-channel production processes are expressed in orange and blue dashed lines.
$y_Q=1$ is fixed for the blue line and the upper limit on Xs$\times$BR$(H^-\to\tau\bar{\nu})^2$ is also shown in red as a comparison.
} 
\end{center}
\end{figure}
%%%%%%%%%%%%%%%%%%%%%%%%%%%%%%%%%%%%%%%%%%%%%%%%%

Based on those considerations, relaxing the upper limit on Xs$\times$BR in bottom flavored di-jet search of Ref.\,\cite{CMS:2018kcg} and Ref.\,\cite{ATLAS:2019itm} by a factor of 2.8 and 3.1 approximately provides the bound on the $bc$ resonance.
We calculate the production cross section allowing up to 2 jets using {\sc\small MadGraph}5\_a{\sc\small MC}@{\sc\small NLO}~\cite{Alwall:2014hca} using {\sc\small NNPDF}2.3 \cite{Ball:2012cx} in the five flavor scheme.
Although the Xs with an additional photon is calculated at the LO,
$H^-$ can also emit the energetic photon and the possible effect of difference kinematic distributions which results in the different acceptance is corrected based on the rapidity cut in Ref.\,\cite{ATLAS:2019itm}.
The resultant constraint with the $bc$ resonance is shown on the $m_H$ versus $y_Q$ plane, Fig.\,\ref{Fig:yQvsmHp}.
The cyan, blue region and purple regions are excluded at 95$\%\,$CL by the bottom flavored di-jet search at $\sqrt{s}=8\,$TeV, flavor inclusive di jet resonance search at $\sqrt{s}=13\,$TeV and
low mass bottom flavored di-jet with a high $p_T$ photon search.
Since the mediator spin dependence in the upper limit on Xs$\times$BR is small \cite{CMS:2018kcg}, we can directly use the given bounds on vector resonance in Refs.\,\cite{CMS:2017dcz,ATLAS:2019itm}.\footnote{The situation is different in a $\tau\nu$ resonance since the chirality of $\tau$ affects the distribution of the hadronic object from $\tau$ decays.}
In this figure other couplings are set to be zero for simplicity.
The constraint from B meson mixings is overlaid in green.
Currently the B meson mixings constraint is stronger than bc resonance ones for $m_H\le325\,$GeV while Run\,1 data gives a stringent upper limit for $m_H\ge325\,$GeV.

The future prospect of the sensitivity  is calculated by assuming the significance grows as $S \propto \sqrt{L}$ based on the observed constraints for Run 2 since those experimental results are consistent with their expectations within $1\sigma$.
The difference between $\sqrt{s}=13\,$TeV and $\sqrt{s}=14\,$TeV is neglected.
The dashed, dotted-dashed, dotted lines correspond to the sensitivity with the integrated luminosity of $139\ifb$, $500\footnote{The value approximately corresponds to the accumulated luminosity at the end of the Run 3 operation.}\ifb$ and $3\iab$, respectively. 
It is noted that the constraint and sensitivity do not rely on the mass difference between heavy neutral scalars.
The HL-LHC is sensitive to $y_Q\sim0.2$ for $m_H=180\,$GeV and $y_Q\sim0.4$ for $m_H=300\,$GeV.

The left handed stau has the same quantum number as that of a charged scalar and a pair of the tauonically decaying scalars contributes to the same signal for $m_{\tilde \chi^0}=0$ where $\tilde \chi^0$ is a neutralino.
As mentioned above in addition to the EW production, a pair of charged Higgs is produced via t-channel topology as shown in the right panel of Fig.\,\ref{Fig:dia_Hp}.
The latter production cross section is proportional to $y_Q^4$ but the former one is independent of the Yukawa couplings.
Although the initial quark species in t-channel processes are charm and bottom,
we see that the Yukawa induced cross section could be comparable when $y_Q$ is of $\mathcal{O}$(1).
As an illustration, we show the t-channel induced production cross section by fixing $y_Q=1$ in dashed blue.
For the comparison the upper limit on Xs$\times$BR$^2$ \cite{CMS:2021woq} is shown in a red solid line.

If BR$(H^-\to\tau\bar{\nu})$ is close to one, the stau bound excludes up to $m_H\simeq340$\,GeV.
However, non zero $y_Q$ reduces BR$(H^-\to\tau\bar{\nu})^2$ rapidly with the help of the color factor in the $bc$ decay mode.
It also contributes to the production cross section, though.
For $m_H> 340$\,GeV, the EW production channel satisfies the current experimental constraint even if BR$(H^-\to\tau\bar{\nu})\simeq 1$ holds.
In this case the parameter set of $|y_Q|\ll1$ and $|y_\tau|\simeq1$ is still allowed.
We will discuss it more quantitatively in the next section.

%%%%%%%%%%%%%%%%%%%%%%%%%%%%%%%%%%%%%%%
\subsection{Current status of the low mass $H^-$ interpretation}
\label{Sec:light_Hp}
%%%%%%%%%%%%%%%%%%%%%%%%%%%%%%%%%%%%%%%
%%%%%%%%%%%%%%%%%%%%%%%%%%%%%%%%%%%%%%%%%%%%%%%%%
\begin{table}[t]
\centering
\newcommand{\bhline}[1]{\noalign{\hrule height #1}}
\renewcommand{\arraystretch}{1.3}
   \scalebox{0.85}{
  \begin{tabular}{c|| c||  c || c || c } 
  Process & Couplings & Mass range  & Number, color & Ref.\\  \hline\hline
  $R_{D^{(*)}}$ & $y_Q \times y_\tau$ & all &\ctext{1}, {{\color[rgb]{0,0.7,0} green}}($1\sigma$) and {{\color[rgb]{0.85,0.85,0} yellow}}($2\sigma$)& \cite{Aoki:2021kgd}\\  \hline 
  $B_c\to\tau\nu$ & $y_Q \times y_\tau$ & all &\ctext{2}, {{\color[rgb]{1,0.62,0.68} light pink}}& \cite{Aebischer:2021ilm}\\  \hline
  B meson mixings & $y_Q$& all &\ctext{3}, {{\color[rgb]{0,1,0} light green}}&\cite{DiLuzio:2019jyq}\\  \hline 
  stau search & $y_\tau$ ($y_Q$) & all &\ctext{4}, {{\color[rgb]{0.85,0,0} red}}&\cite{CMS:2021woq}\\  \hline 
  2b & $y_Q$ ($y_\tau$) &$m_H\ge325\,$GeV&\ctext{5}, {{\color[rgb]{0,1,1} cyan}}&\cite{CMS:2018kcg}\\  \hline
  2j &$y_Q$ ($y_\tau$)&$m_H\le300\,$GeV&\ctext{6}, {{\color[rgb]{0,0,1} blue}}&\cite{CMS:2017dcz}\\  \hline 
  2b+$\gamma$ &$y_Q$ ($y_\tau$)&$m_H\ge225\,$GeV&\ctext{7}, {{\color[rgb]{0.4,0,0.4} purple}}&\cite{ATLAS:2019itm}\\  \hline 
  $\tau\nu$ (Run\,1)&$y_Q \times y_\tau$& $m_H\ge300\,$GeV&\ctext{8}, {{\color[rgb]{1,0.4,0.1} orange}}&\cite{CMS:2015hmx}\\  \hline 
  $\tau\nu$ (Run\,2)&$y_Q \times y_\tau$& $m_H\ge400\,$GeV&\ctext{9}, {{\color[rgb]{0.4,0.4,0.4} grey}}&\cite{Sirunyan:2018lbg}\\  \hline 
   \end{tabular}
   }
    \caption{\label{Tab:cosntraint} 
    The list of the relevant constraint, relevant couplings and mass range, number in the figure and corresponding colors are summarized.
    The current LHC bound is expressed in solid line and future prospect with 139fb$^{-1}$, 500fb$^{-1}$ and 3ab$^{-1}$ of the data is shown in dashed, dotted-dashed, dotted lines in the same color.
}
\end{table}
%%%%%%%%%%%%%%%%%%%%%%%%%%%%%%%%%%%%%%%%%%%%%%%%%
Based on those results we discuss the current status of the charged scalar interpretation of the $R_{D^{(*)}}$ anomaly in the light mass window.
Depending on the heavy scalar masses different constraints are relevant and thus we show the result in 10 mass points, $m_H=180,\,200,\,225,\,250,\,275,\,300,\,325,\,350,\,375,\,400\,$GeV as a demonstration. 
The $R_{D^{(*)}}$ favored region and various constraints in the $y_Q$ versus $y_\tau$ plane are shown by fixing the mass in Fig.\,\ref{Fig:coupling}.
The value of the fixed mass is shown in upper right of each figure.
The B meson mixing constraints and collider constraints do not depend on the imaginary phase of the Yukawa couplings.
On the other hand the phase affects the favored range of the Yukawa coupling for the $R_{D^{(*)}}$ anomaly and constraint from $B_c\to\tau\nu$.
We will discuss them bellow.
We assign the number on the each constraint based on  Tab.\,\ref{Tab:cosntraint} because that a number of the constraints and prospects is large and figure looks busy when we describe them all on the plot.

%%%%%%%%%%%%%%%%%%%%%%%%%%%%%%%%%%%%%%%%%%%%%%%%%
\begin{figure}[p]
\begin{center}
\includegraphics[scale=0.315]{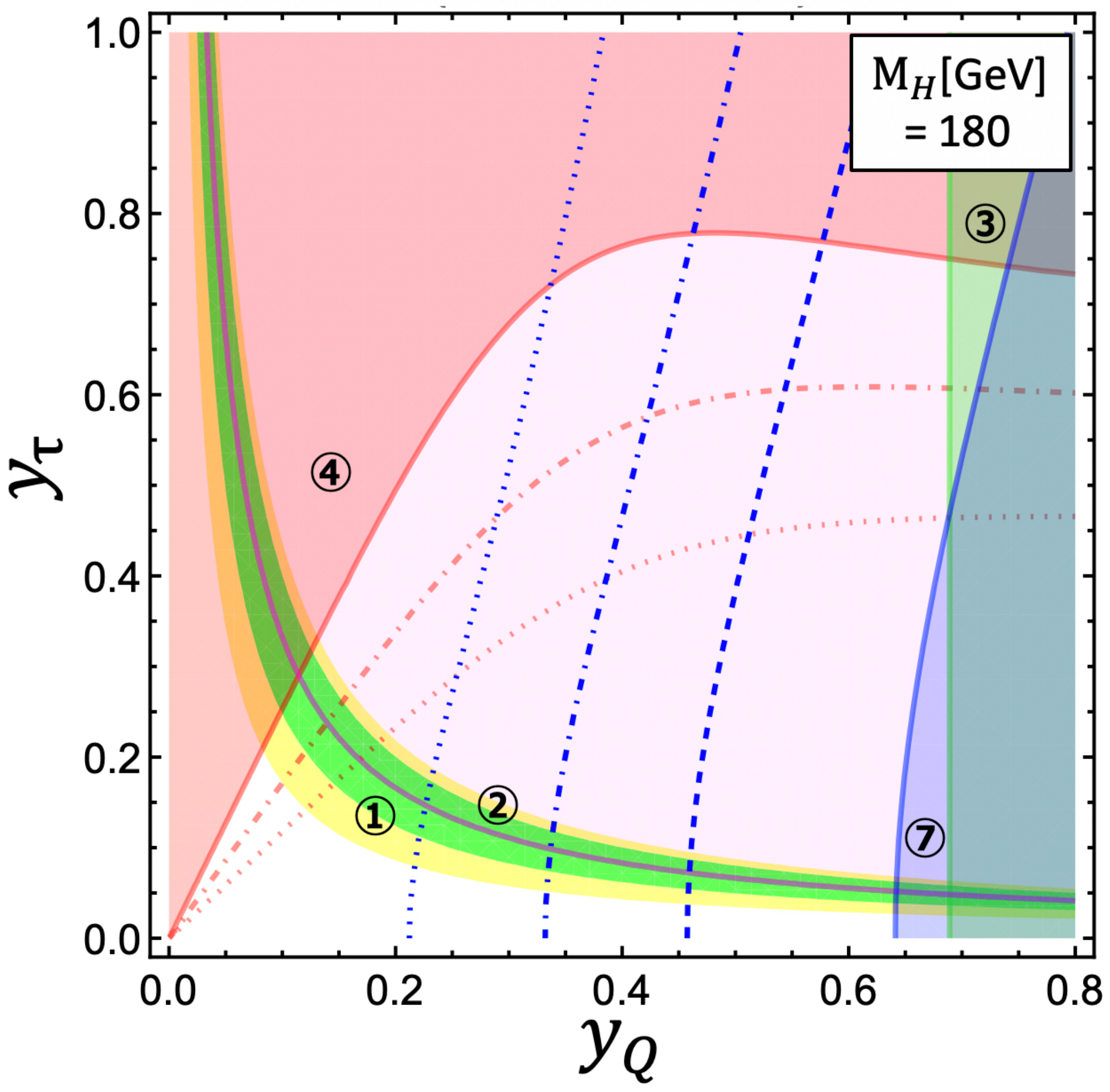}
\includegraphics[scale=0.315]{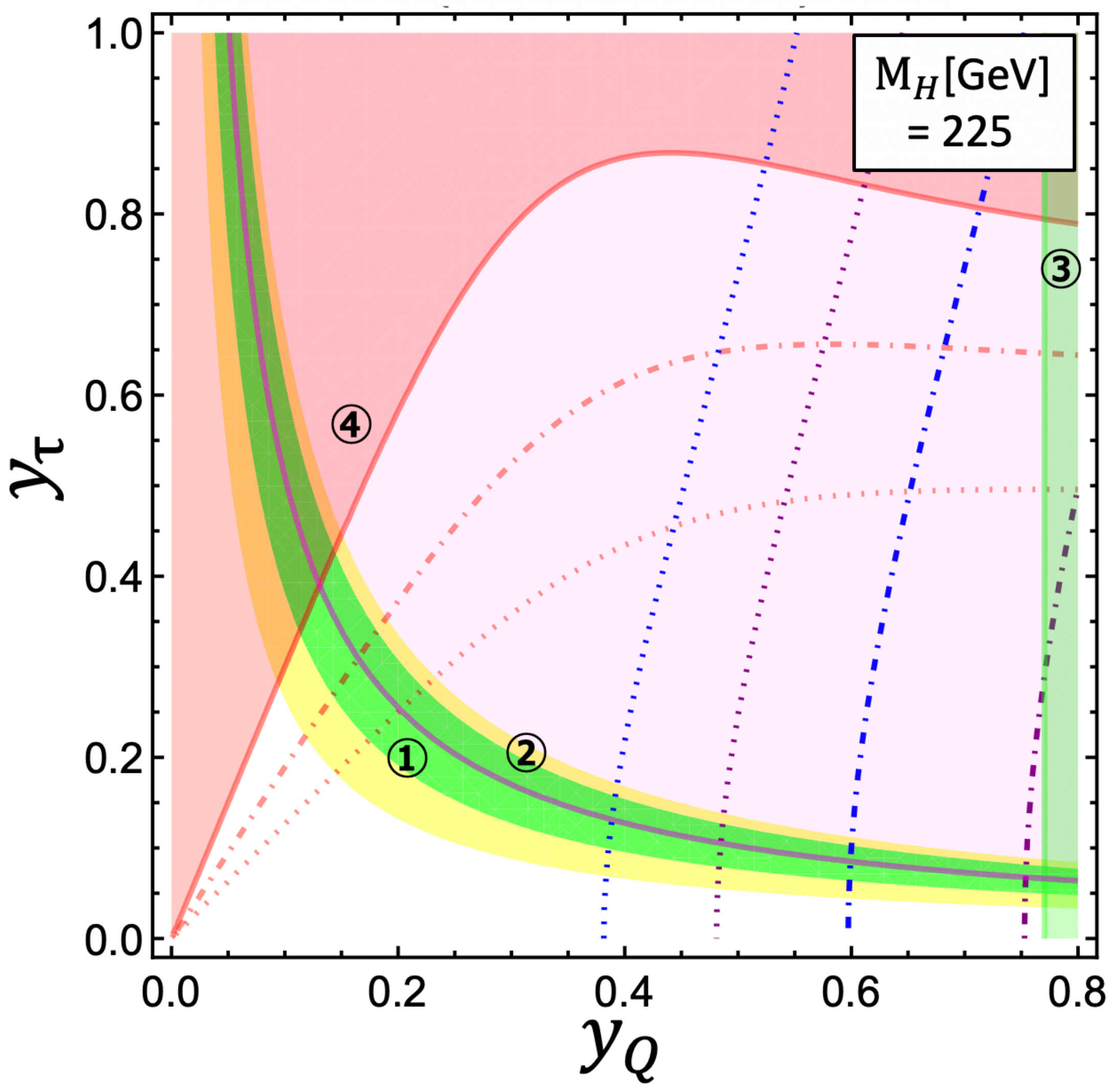}
\includegraphics[scale=0.315]{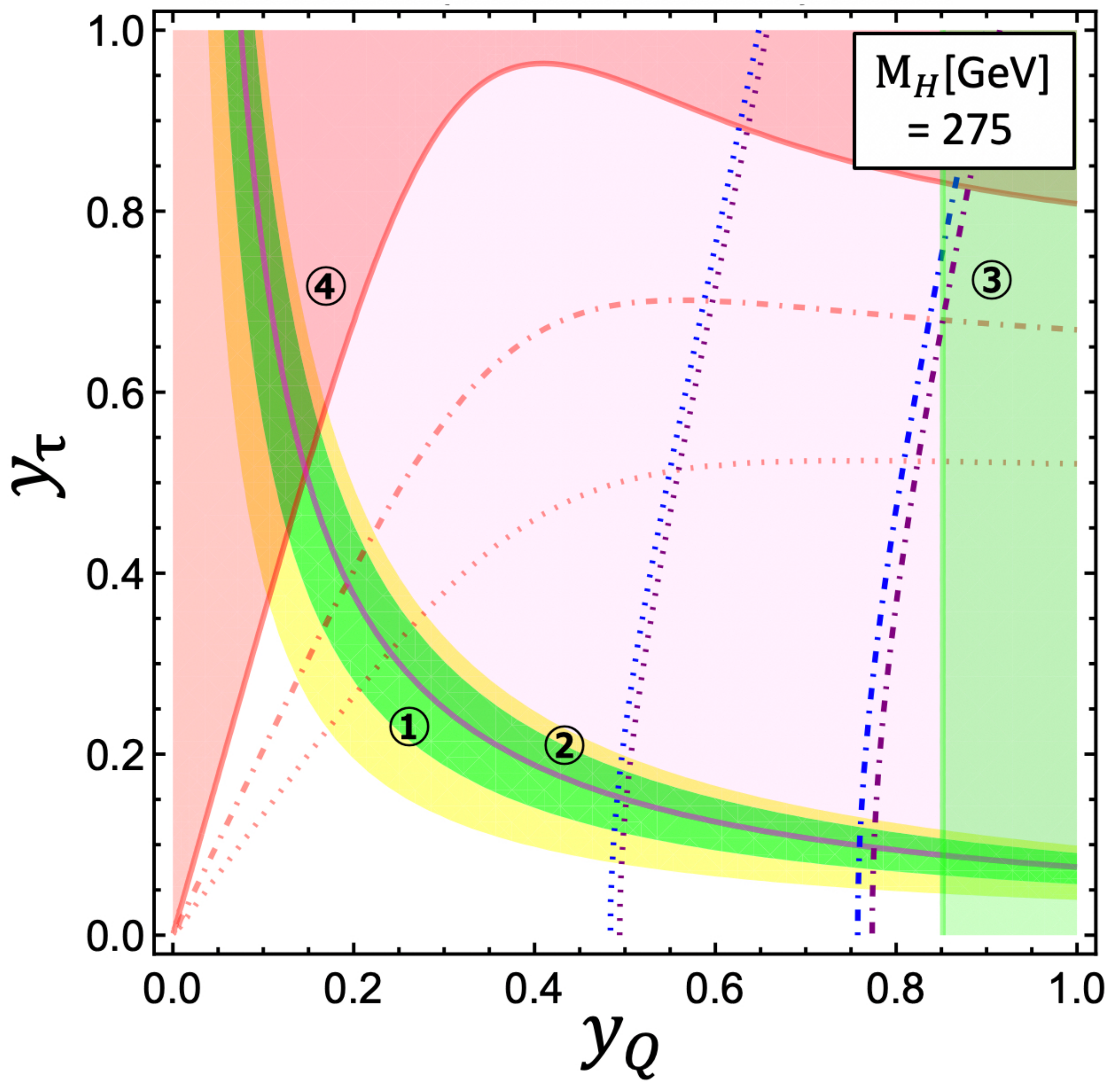}
\includegraphics[scale=0.315]{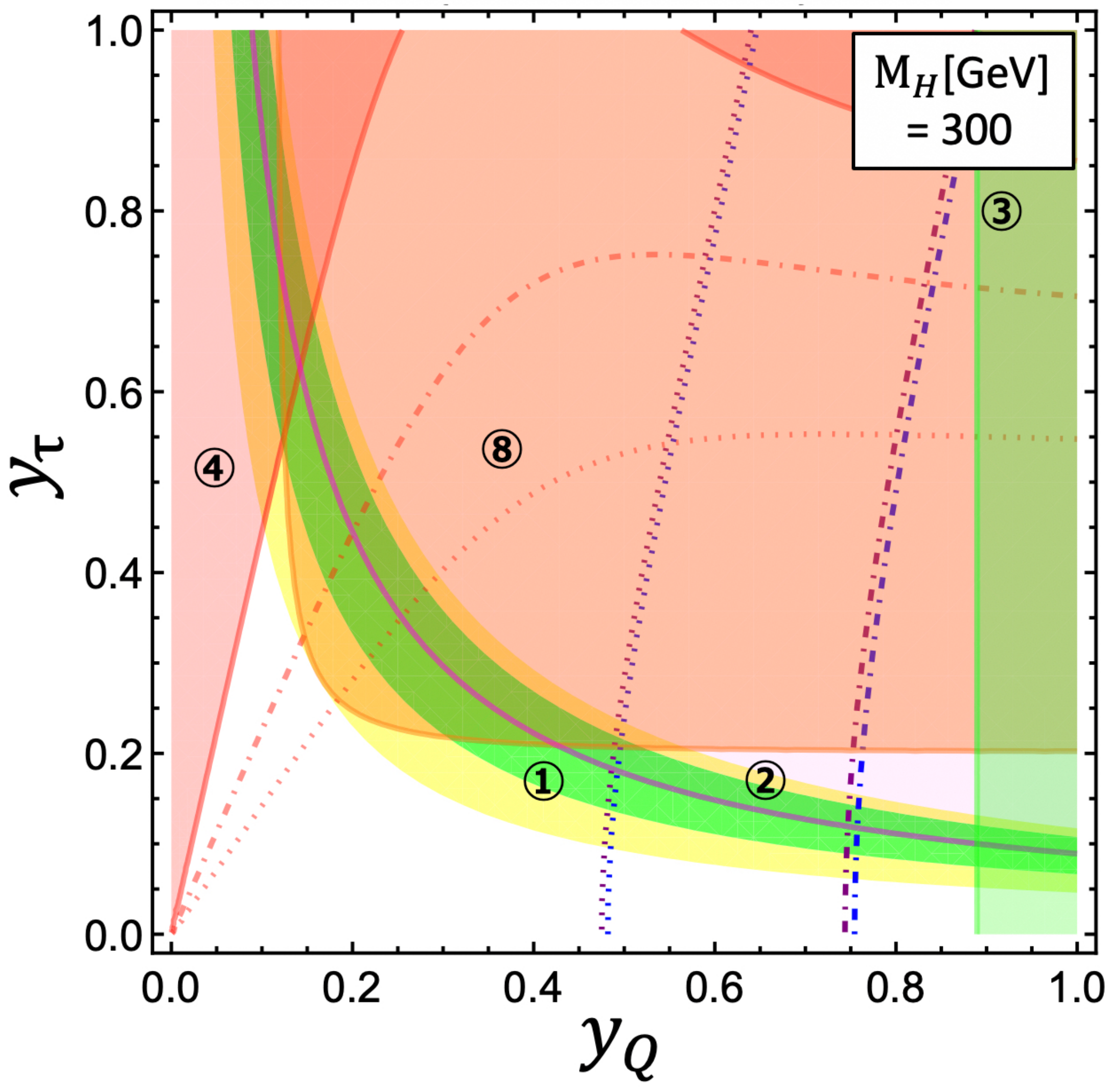}
\includegraphics[scale=0.315]{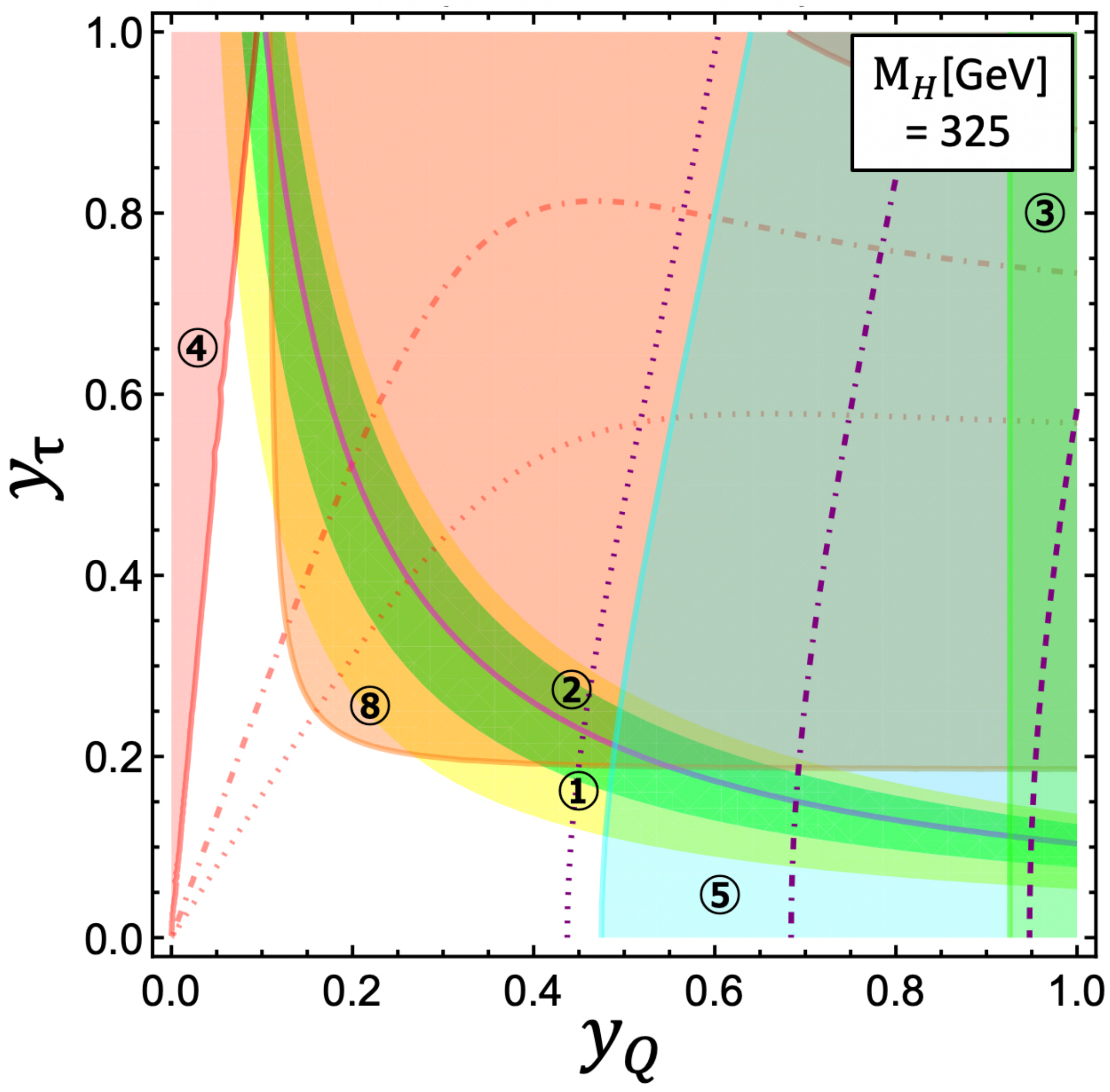}
\includegraphics[scale=0.315]{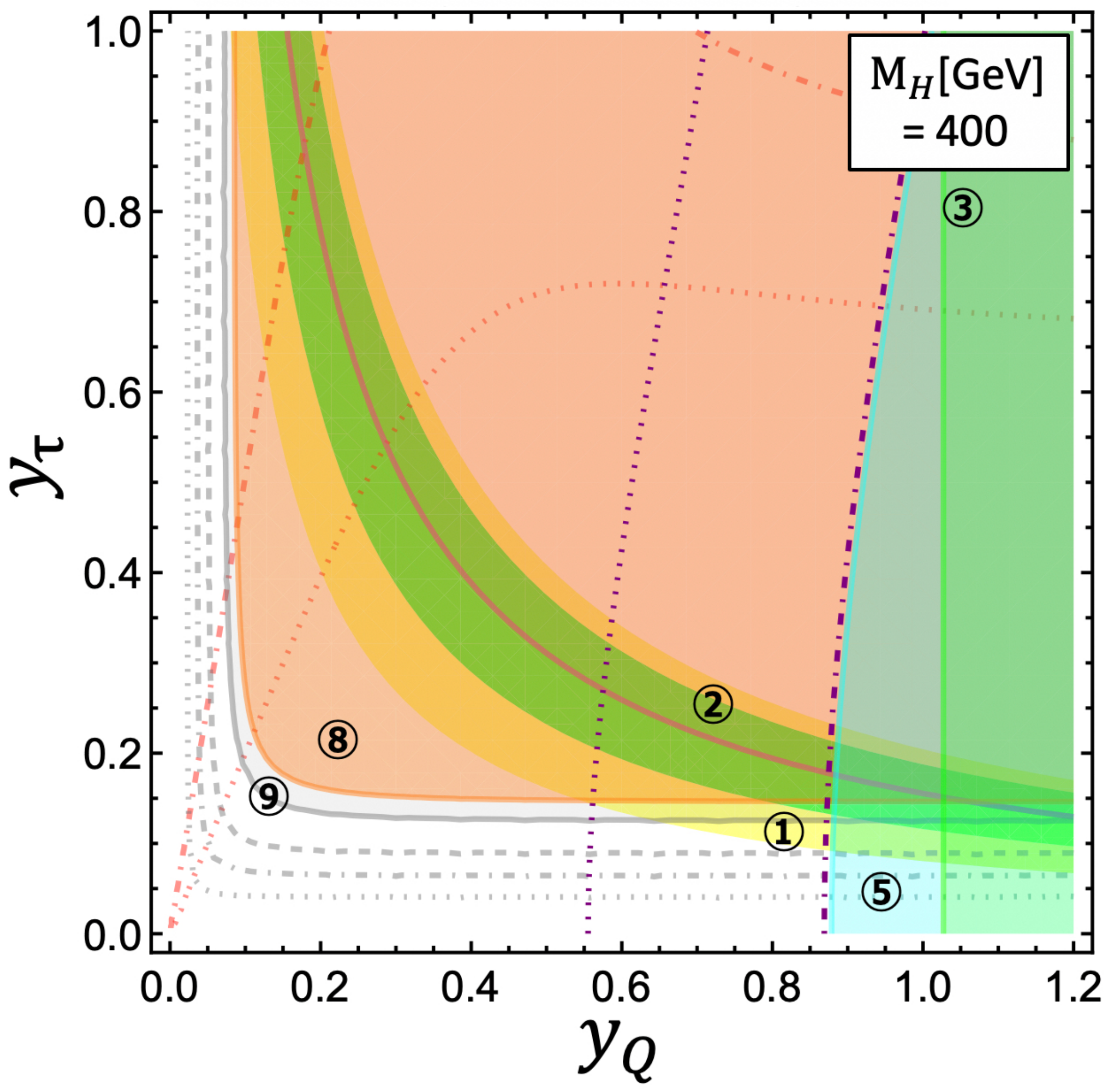}
\caption{
\label{Fig:coupling} 
Flavor and collider constraints on the coupling plane. 
The mass fixed in each plane is shown in upper right.
The circled numbers express the relevant observables and processes defined in Tab.\,\ref{Tab:cosntraint}. 
Solid lines show the current constraint while dashed, dotted-dashed, dotted lines correspond to the projected sensitivity with $139\,\ifb,\,500\,\ifb,\,3\,\iab$.
Figures with $m_H=200,\,250,\,350,$ and $375$\,GeV are put in Appendix\,\ref{Sec:App_other_mass} because of the space.
} 
\end{center}
\end{figure}
%%%%%%%%%%%%%%%%%%%%%%%%%%%%%%%%%%%%%%%%%%%%%%%%%

When we calculate $C_{S_L}$ at the $m_b$ scale, the renormalization group running corrections are taken into account \cite{Alonso:2013hga, Jenkins:2013wua,Gonzalez-Alonso:2017iyc,Feruglio:2018fxo}, which was not considered in Ref.\,\cite{Iguro:2018qzf}.
This correction is found to be important to judge the availability since the correction amplifies the scalar contribution at $m_b$.
For instance the relation, $C_{S_L}(m_b)\simeq 1.6\,C_{S_L}(m_H)$ holds when $m_H=300\,$GeV is assumed.
The range of the required absolute value of $C_{S_L}$ at the $m_b$ scale is $[0.84,\,1.36]$ for $1\sigma$ and $[0.58,\,1.45]$ for $2\sigma$.

The bands for $R_{D^{(*)}}$ favored coupling products are calculated by fitting the phase to minimize  $\chi^2$\footnote{Only $R_D$ and $R_{D^*}$ are considered in calculating $\chi^2$.} and shown in green (1$\sigma$) and yellow (2$\sigma$).
The upper limit on the coupling product from BR$(B_c\to\tau\nu)\le63\%$ shown in pink is obtained so that $\chi^2$ is minimized with respecting the bound.
The light green region is constrained by B meson mixings.
The constraint from the current stau search is shown in red with the corresponding prospect with 500,\,3000\,fb$^{-1}$ of the data.
Cyan, blue and purple shaded regions are excluded by  di-jet searches and the same coloring scheme is used as in Fig.\,\ref{Fig:yQvsmHp} (left).
The constraints from $\tau\nu$ resonance search based on 20\,fb$^{-1}$ of the data at Run\,1 and 36\,fb$^{-1}$ of the data at Run\,2 are shown in orange and grey, respectively.
The HL-LHC prospect is calculated assuming 139,\,500,\,3000\,fb$^{-1}$ of the data and shown in dashed, dotted-dashed, dotted lines.
\footnote{The CMS result at $\sqrt{s}=13\,$TeV has the deficit in the number of observed events in large $m_T$ region and it results in the stringent constraint on leptoquark models. 
However the result in the low mass region is consistent with their expectation.
Therefore we rescaled the observed constraint in Ref.\cite{Iguro:2018fni} to obtain the future sensitivity.} 

%%%%%%%%%%%%%%% Interpretation %%%%%%%%%%%%%%%%%
As is shown in Fig.\,\ref{Fig:current_EFT} the $B_c\to\tau\nu$ constraint can not exclude all of the $1\sigma$ explanations and it is observed that various constraints are very complementary, see Fig.\,\ref{Fig:coupling}.
Depending on to which coupling they are sensitive various constraints are roughly categorized into three:
\begin{enumerate}
    \item { Observable sensitive to $y_Q$ e.g. $\Delta M_{B_s}$ and $bc$ resonance.}
    \item { One sensitive to the coupling product $y_Q\times y_\tau$ e.g. $R_{D^{(*)}}$, $B_c\to\tau\nu$ and $\tau\nu$ resonance.}
    \item { The search sensitive to the balance of $y_Q$ and $y_\tau$ e.g. stau search.}
    \end{enumerate}
The observables in the category 1 probe the scenario from right to left on the plane and the one in the category 2 tests from upper right to the origin.
Although the stau search which belongs to the category 3 probes the parameters region with $y_\tau\gg y_Q$, it also depends on the mass assumption.

For the $m_H=180$\,GeV case (upper left), the stau constraint excludes the large $y_\tau$ scenario since the EW production Xs is larger than the current experimental bound.
However, once $y_Q$ is getting larger, the constraint gets weaker.
The larger $y_Q$ region is excluded by the flavor inclusive low mass resonance search.
As a result we observe that there is the available range of $y_Q$ which is not accessible information only with $B_c\to\tau\nu$.
The HL-LHC sensitivity shown in dashed, dotted-dashed, dotted lines shows that the wide range of the parameter space can be probed.,
However, it is not possible to test the all parameter space even at the end of the HL-LHC.
The result for $m_H=225$\,GeV (upper right) and $m_H=275$\,GeV (middle left) scenarios are similar to the $m_H=180$\,GeV one but current di-jet constraint is less stringent and B meson mixing gives the upper bound on $y_Q$.
However, the future data can probe the wide range of the parameter space.
Since $\tau\nu$ resonance searches are not available for $m_H\le300$\,GeV, not all of the currently favored region can be covered.

On the other hand, once the $\tau\nu$ resonance result becomes available the situation changes.
The combination with the current stau bound can constrain the solution with $y_\tau\ge y_Q$ when $m_H=300$\,GeV is taken and the projected sensitivity at the HL-LHC will greatly cover the $1\sigma$ range (middle right).
As for the $m_H=325$\,GeV case (lower left), the bottom flavored di-jet search at $\sqrt{s}=8\,$TeV is stringent and already covers most of the solution with $y_\tau\le y_Q$.
However, the parameter still exists in $y_\tau\ge y_Q$, it will be probed in near future by the stau search.
$\tau\nu$ resonance searches are found to be powerful for $m_H=400\,$GeV (lower right) and combining di-jet search allows us to exclude the $1\sigma$ solution.
Furthermore $2\sigma$ solutions will be also probed with the HL-LHC data.
Therefore by combining various constraints we can cover the vast of interesting parameter region and lowering the threshold for $\tau\nu$ resonance searches is highly desired to probe the all of parameter space in the light mass window.

%%%%%%%%%%%%%%%%%%%%%%%%%%%%%%%%%%%%%%%%%%%%%%%%%%%
\section{\boldmath Conclusions and discussion}
\label{Sec:summary}
%%%%%%%%%%%%%%%%%%%%%%%%%%%%%%%%%%%%%%%%%%%%%%%%%%
%
The experimental results from B-factories have indicated a discrepancy between the measurement and the SM predictions in $R_{D^{(*)}}$.
It has been known that $B_c\to\tau\nu$ stringently constrains the charged scalar interpretation of the anomaly, however, the recent re-evaluation showed that the current conservative bound is BR$(B_c\to\tau\nu)\le 63\%$ mainly due to the large charm mass uncertainty.
We pointed out that it is still possible to explain the 1$\sigma$ region within a G2HDM if we apply this bound.
Furthermore the scalar contribution can enhance $F_L^{D^*}$.
In order to generate large deviations, the charged Higgs mass needs to be less than O(1) TeV even is its Yukawa couplings are of $\mathcal{O}$(1).
Therefore it is natural to search the new particles at the LHC. 
The previous study found the heavy $\tau\nu$ resonance search at the CMS with 36 fb$^{-1}$ of the data gives more stringent constraint for $m_H \ge 400$\,GeV and excludes the interpretation.
On the other hand the experimental data at Run\,2 is not available for $m_H \le 400$\,GeV because the search originally looks for the heavy $W^\prime$ in the sequential standard model and the huge W boson BG exists in the light region. 
The Run\,1 result, however, was already available, its less stringent constraint was not used in the previous paper since we wanted to set the bound on the heavier scenario. 

In this work, we focused on the light mass region $180$\,GeV$\le m_H \le 400$\,GeV and studied the LHC sensitivity for the light charged Higgs interpretation of the $R_{D^{(*)}}$ anomaly.
The constraints from the stau search, low mass flavor inclusive and bottom flavored di-jet searches, $\tau \nu$ resonance searches, B meson mixings are derived.
It was found that those constraints are complementary to constrain the available parameter space.
For instance we found the $m_H=325\,$GeV scenario is nearly covered by combining constraints.
The future sensitivity is also shown and most of the parameter space for $m_H\ge300\,$GeV will be covered by extending the existent searches.

In this work the $bb$ resonance constraint is rescaled to obtain the bound for the $bc$ resonance by considering the difference in the tagging efficiencies.
The requirement of the higher QCD jet rejection rate in the bottom tagging tends to suppress the mistag rate of $\slashed{c} \to b$.
In the coming high luminosity era, the requirement of high purity in a bottom tagging would be good to improve the $bb$ resonance sensitivity.
However, it does not always maximize the sensitivity to the $bc$ resonance as long as the rescaling procedure is applied.
If one requires two b-tagged jets passing the tight working point the conversion factor is more than 10 which was estimated to be 2.8 and 3.1 for Ref.\,\cite{CMS:2018kcg} and Ref.\,\cite{ATLAS:2019itm}, and thus the sensitivity to $bc$ resonances gets worse.
The more careful experimental study for the $bc$ resonance would be interesting.

It is inferred that the requirement of an additional heavy flavored jet in bc resonance search would improve the sensitivity to the charged scalar since there is the PDF enhanced $gc\to bH^-\to b \bar{b}c$ process.
An estimation of the size of QCD jet BG is difficult without the data driven technique and the experimental analysis is also desired.
For instance Ref.\,\cite{ATLAS:2021suo} searched for bottom flavored di-jet resonances with additional b-tagged jets, however, they looked for $m_{bb}>\mathcal{O}(1)$\,TeV. 

It could be important to point out that the bound and prospect of $y_Q$ in Fig.\,\ref{Fig:yQvsmHp} on the left also would have a great impact on electroweak baryogenesis driven by complex Yukawa couplings \cite{Fuyuto:2017ewj} and spontaneous CP violating potential within a G2HDM \cite{Nierste:2019fbx}.

In the light mass region a requirement of an additional b-tagged jet in $\tau\nu$ resonance search can suppress the SMBG and improve the signal sensitivity which has not been performed in the experiments.
From the result obtained in Refs.\,\cite{Altmannshofer:2017poe,Iguro:2017ysu,Abdullah:2018ets,Marzocca:2020ueu,Iguro:2020keo,Endo:2021lhi}, it is possible to infer that this additional b-tagging technique and selecting negatively charged 
$\tau$ events are also effective to probe the low mass window.
Revision of this problem is my future work \cite{Blanke:2022pjy}.
In this work the collider phenomenology of neutral scalars is not discussed.
The single neutral scalar production $gc\to t\phi$ with a subsequent decay of $\phi\to\tau\bar{\tau}$ would be useful since the SMBG is expected to be not huge \cite{ATLAS:2020bhu}.

%%%%%%%%%%%%%%%%%%%%%%%%%%%%%%%%%%%
%%%%%%%%%%%%%%%%%%%%%%%%%%%%%%%%%%%

%%%%%%%%%%%%%%%%%%%%%%%%%%%%%%%%%%%%%%%%%%%%%%%%%%%
\section*{\boldmath Acknowledgement}
\label{Sec:acknowledgement}
%%%%%%%%%%%%%%%%%%%%%%%%%%%%%%%%%%%%%%%%%%%%%%%%%%%
%%%
I would like to thank Joaquim Matias, Ryoutaro Watanabe, Hiroyasu Yonaha and Teppei Kitahara for encouraging this project.
I wish to appreciate Kazuhiro Tobe
for careful reading of the manuscript and comments that helped to improve the paper.
I also thank Yuta Takahashi, Hantian Zhang, Ulrich Nierste, Monika Blanke for the fruitful discussion. 
%%%
I enjoy the support from the Japan Society for the Promotion of Science (JSPS) Core-to-Core Program, No.JPJSCCA20200002 and the Deutsche Forschungsgemeinschaft (DFG, German Research Foundation) under grant 396021762-TRR\,257.
%%%
%%%%%%%%%%%%%%%%%%%%%%%%%%%%%%%%%%%%%%%%%%%%%%%%%%%

%%%%%%%%%%%%%%%%%%%%%%%%%%%%%%%%%%%%%%%
\appendix
%%%%%%%%%%%%%%%%%%%%%%%%%%%%%%%%%%%%%%%
\section{$B$ meson mixings}
\label{Sec:App_flavor}
%%%%%%%%%%%%%%%%%%%%%%%%%%%%%%%%%%%%%%%
In this appendix, the $H^-$ contribution to $B_{s(d)}$--$\Bb_{s(d)}$ mixing is discussed.
Tree level neutral scalar contribution is absent when $y_{Q_d}=0$ is assumed in Eq.\,(\ref{Eq:G2HDM}). 
The 1-loop $H^-$ box contribution to $\Delta M_{B_s}$ is given as~\cite{Iguro:2017ysu}
\beq
\frac{\Delta M_{B_s}}{\Delta M_{B_s}^{\rm SM}}
= \left| 1 + \frac{C_{B_s}^{\rm NP} (M_\text{W})}{C_{B_s}^{\rm SM} (M_\text{W})}\right|\,,
\eeq
with 
\begin{align}
C_{B_s}^{\rm NP} (M_\text{W})&\simeq \left(\frac{\alpha_s (M_{\rm NP})}{\alpha_s (M_W)}\right)^{\frac{2}{7}} C_{B_s}^{\rm NP} (M_\text{NP}),~~~
C_{B_s}^{\rm SM} = -2.35 \frac{\left( V_{tb}V_{ts}^\ast G_F M_W\right)^2}{4 \pi^2}\,,\\ 
C_{B_s}^{\rm NP} (M_\text{NP})&\simeq\frac{(V_{tb}V_{ts}^*)^2|y_{Q_L}|^4}{128\pi^2m_{H^-}^2}G_1\biggl(\frac{m_c^2}{m_{H^-}^2}\biggl),~~~
G_1(x)=\frac{-1+x^2-2x\,\text{log}[x]}{(1-x)^3}
\end{align}
and
\beq
\mathcal{H_{\rm eff}}= -C_{B_s} \left(\overline{s}\gamma^{\mu} P_L b\right)
\left(\overline{s}\gamma_{\mu} P_L b\right)\,.
\eeq
Here, the WC $C_1$, is evaluated at the electroweak scale, and the 1-loop QCD correction from the RG evolution~\cite{Bagger:1997gg} is considered.
The formula for 1-loop box is taken from
Appendix C of Ref.\,\cite{Iguro:2017ysu} with replacing $\rho_{tc}\to y_{Q}$ to change the notations.
We newly consider the RG running effect.
Following Ref.\,\cite{DiLuzio:2019jyq} we impose 
 $0.88<{\Delta M_{B_s}}/{\Delta M_{B_s}^{\rm SM}} <1.10$ in the numerical analysis.
The corresponding relations for $B_{d}$--$\Bb_{d}$ mixing cen be obtained by replacing the indices.
The constraint from $\Delta M_{B_d}$ is similar to $\Delta M_{B_s}$ and is omitted.
It is noted that the leading box contribution is proportional to $|y_Q|^4$.
$\Delta M_{B_s}/\Delta M_{B_d}$ is the same as SM and can not be helpful to constrain the model.

%%%%%%%%%%%%%%%%%%%%%%%%%%%%%%%%%%%%%
%%%%%%%%%%%%%%%%%%%%%%%%%%%%%%%%%%%%%
\section{Additional figures}
\label{Sec:App_other_mass}
We show the BR($H^-\to\tau\bar{\nu}$) and width to mass ratio on the $y_Q$ versus $y_\tau$ plane in Fig.\,\ref{Fig:App_BR}.
The blue solid lines express the BR of $\tau\nu$ mode and red dashed lines present the width to mass ratio.
The masses of the final state are neglected since we focus on the light mass window defined in Eq.\,(\ref{Eq:window}).

%%%%%%%%%%%%%%%%%%%%%%%%%%%%%%%%%%%%%%%%%%%%%%%%%
\begin{figure}[h]
\begin{center}
\includegraphics[scale=0.25]{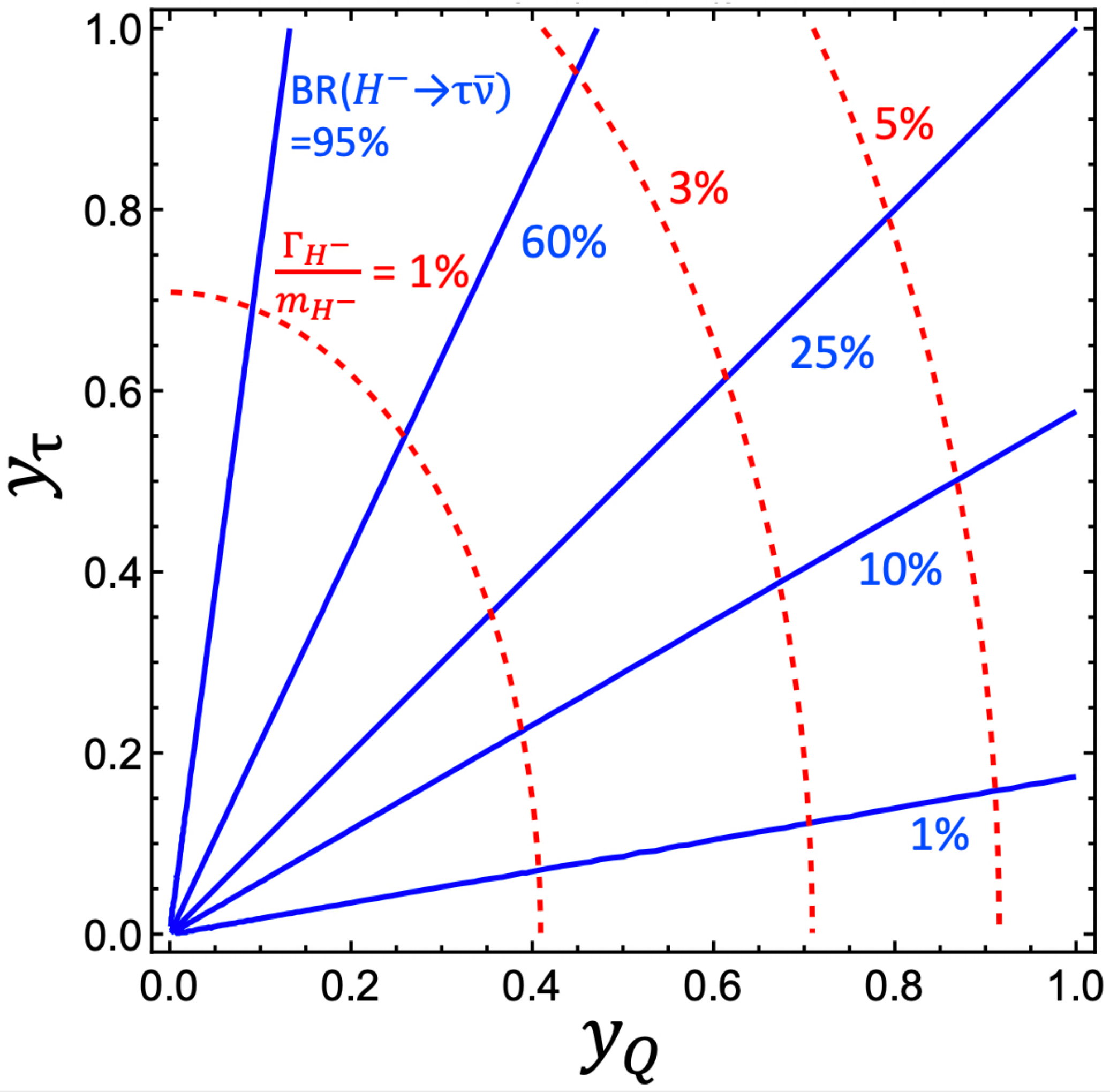}
\caption{The BR($H^-\to\tau\bar{\nu}$) and width to mass ratio on the $y_Q$ versus $y_\tau$ plane are shown.
\label{Fig:App_BR}
} 
\end{center}
\end{figure}
%%%%%%%%%%%%%%%%%%%%%%%%%%%%%%%%%%%%%%%%%%%%%%%%%
%%%%%%%%%%%%%%%%%%%%%%%%%%%%%%%%%%%%%%%%%%%%%%%%%
\begin{figure}[h]
\begin{center}
\includegraphics[scale=0.28]{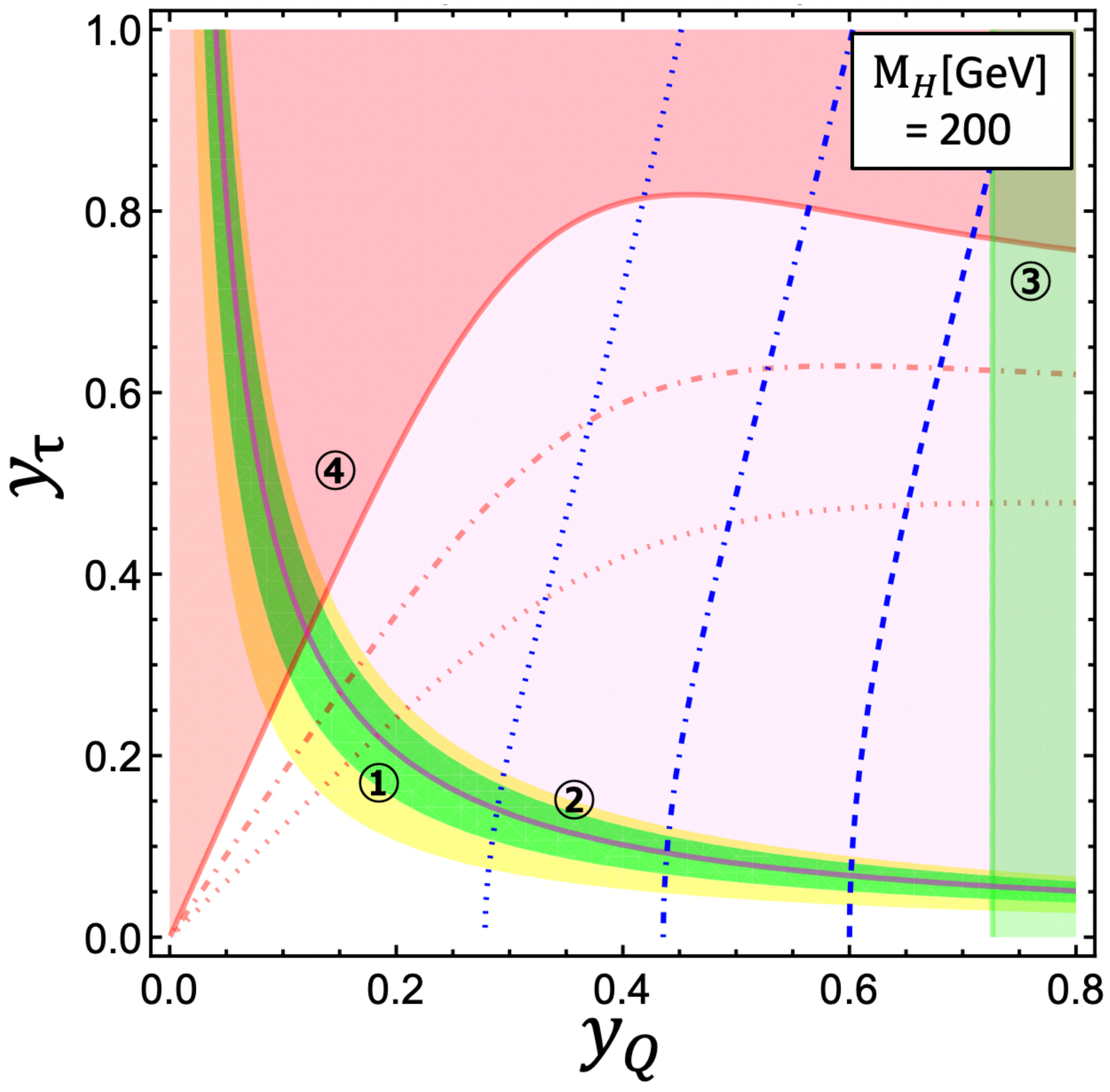}
\includegraphics[scale=0.28]{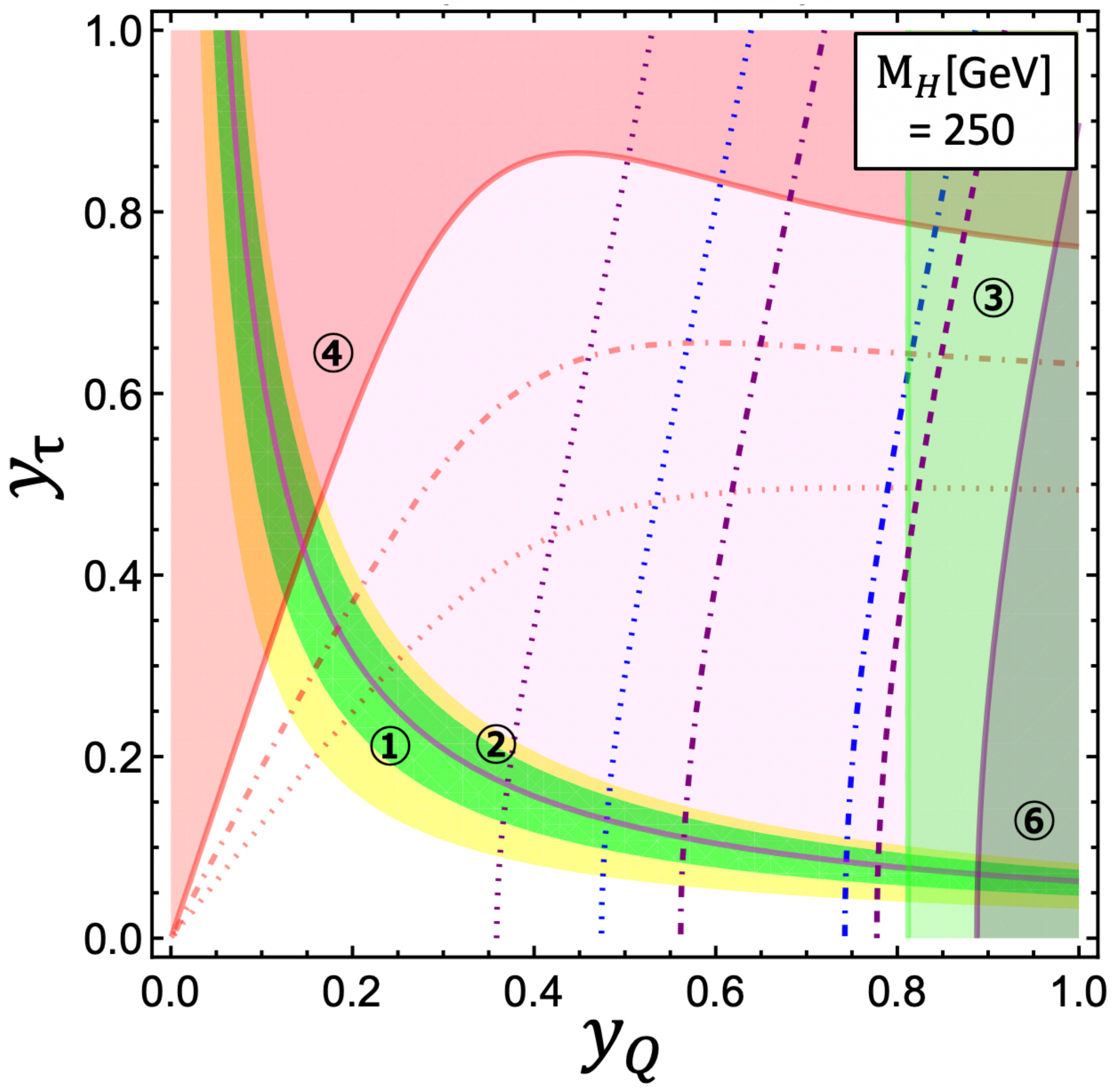}
\includegraphics[scale=0.28]{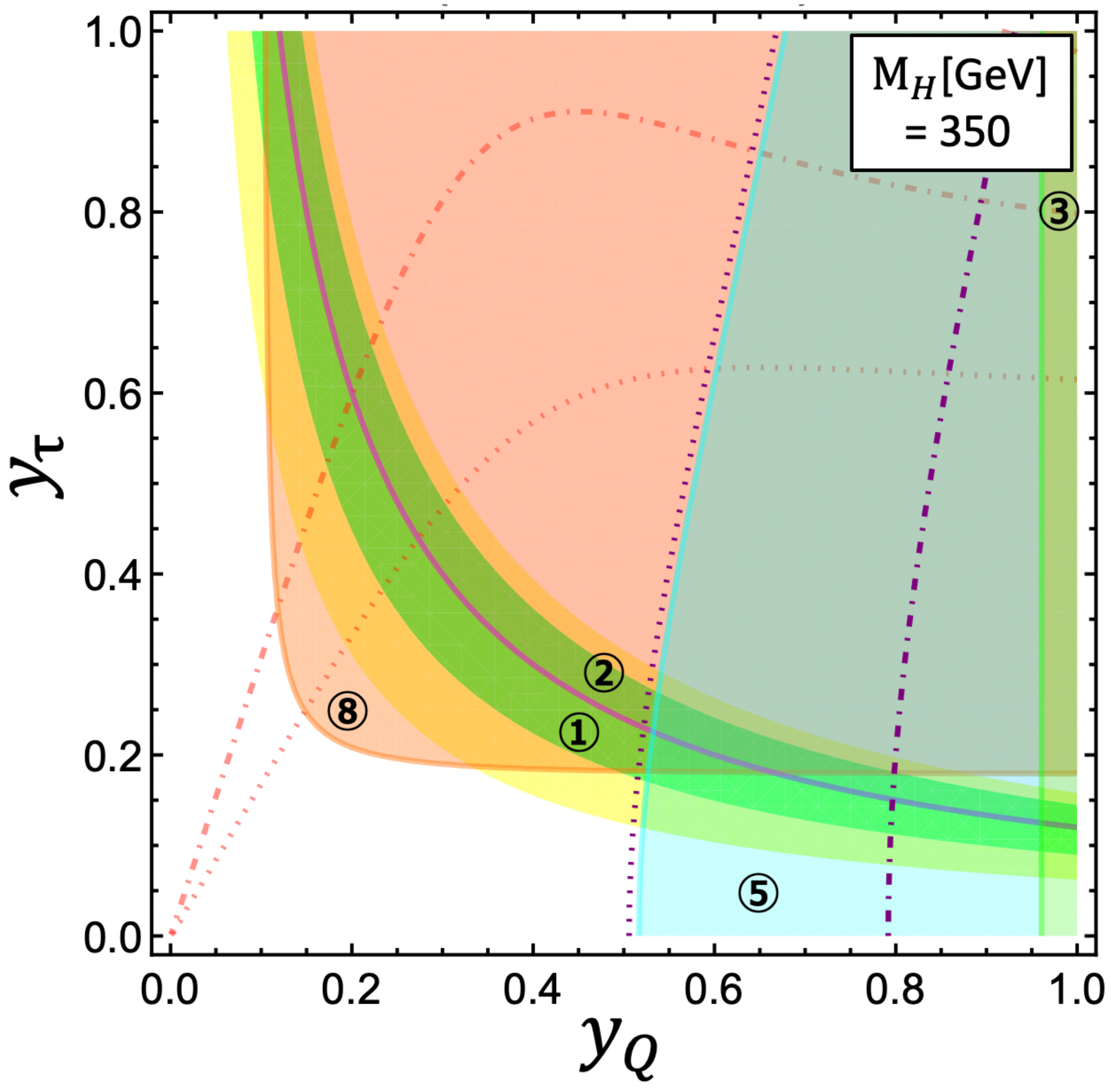}
\includegraphics[scale=0.28]{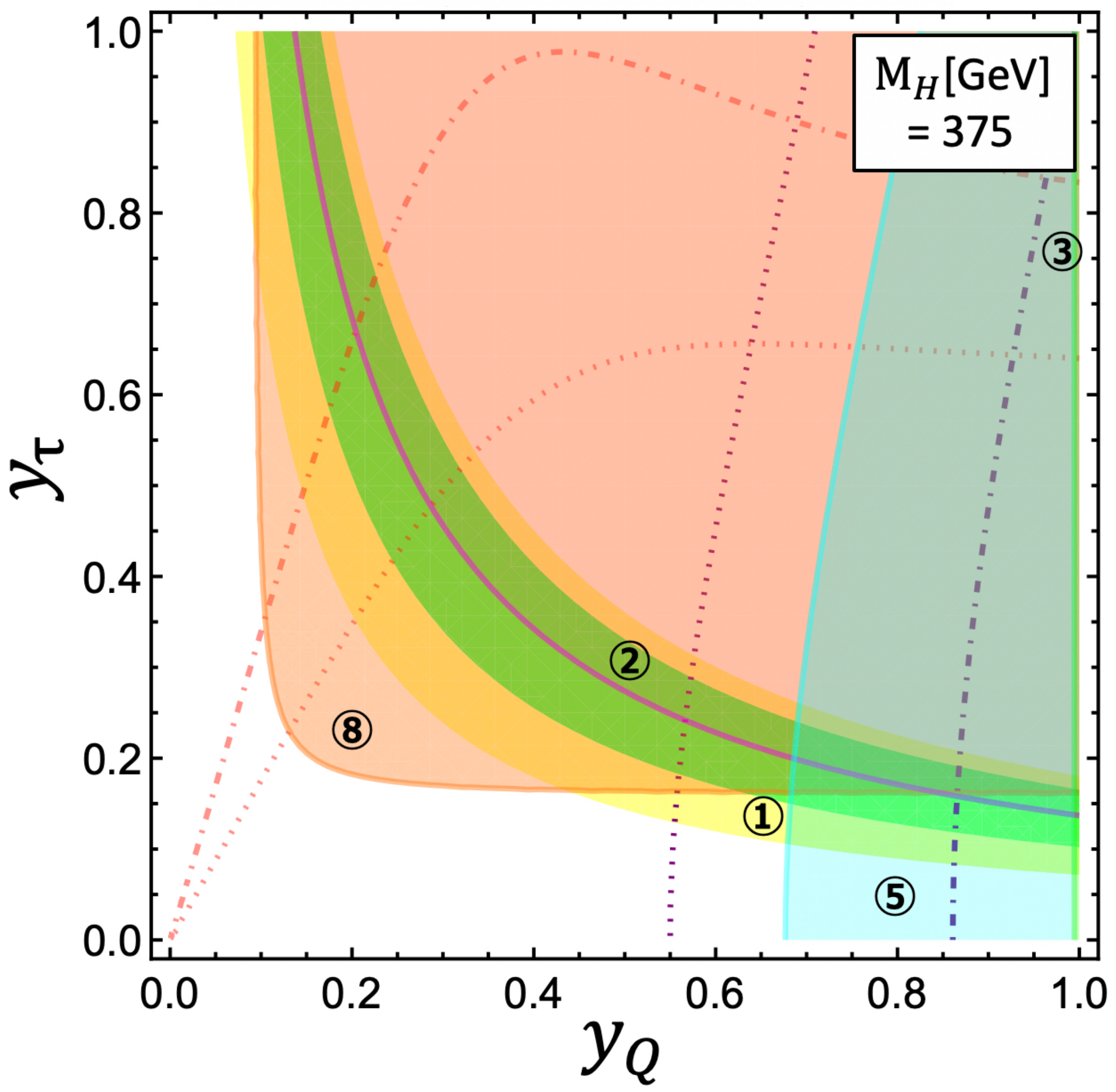}
\caption{
\label{Fig:App_coupling}
Flavor and collider constraints on the coupling plane. 
The mass fixed in each plane is shown in upper right.
The circled numbers express the relevant observables and processes defined in Tab.\,\ref{Tab:cosntraint}. 
} 
\end{center}
\end{figure}
%%%%%%%%%%%%%%%%%%%%%%%%%%%%%%%%%%%%%%%%%%%%%%%%%
In Fig.\,\ref{Fig:App_coupling} the result for $m_H=200,\,275,\,350,$ and $350$\,GeV which is not included in Fig.\,\ref{Fig:coupling} is shown.
The color scheme is the same and readers are referred to Tab.\,\ref{Tab:cosntraint}.

%%%%%%%%%%%%%%%%%%%%%%%%%%%%%%%%%%%%%%%
%%%%%%%%%%%%%%%%%%%%%%%%%%%%%%%%%%%%%
%%%%%%%%%%%%%%%%%%%%%%%%%%%%%%%%%%%%%

\clearpage
\bibliographystyle{utphys28mod}

\bibliography{ref}

\end{document}
%%%%%%%%%%%%%%%%%%%%%%%%%%%%%%%%%%%%%
%%%%%%%%%%%%%%%%%%%%%%%%%%%%%%%%%%%%%